\documentclass[numberedappendix]{emulateapj}
\usepackage{array,wasysym}

\newcommand{\kfive}{{Kepler}-5{\it b}}
\newcommand{\kfour}{{Kepler}-4{\it b}}
\newcommand{\kic}{{KIC 1255{\it b}}}
\newcommand{\bone}{[1, 2, 3, 5]}
\newcommand{\btwo}{[1, 2, 0, 3, 0, 0, 5, 0, 0, 0, 0]}
\newcommand{\kepler}{{\it Kepler}}

\newcommand{\wtf}{KIC 8462852}
\newcolumntype{L}[1]{>{\raggedright\let\newline\\\arraybackslash\hspace{0pt}}p{#1}}

\slugcomment{Accepted to ApJ}

\shorttitle{Signatures of Transiting Megastructures}
\shortauthors{Wright et al.}

\begin{document}


\title{The \^G Search for Extraterrestrial Civilizations with Large Energy Supplies.\\ IV. The Signatures and Information Content of Transiting Megastructures}

\author{Jason T.\ Wright\altaffilmark{1}, Kimberly M.\ S.\ Cartier, Ming Zhao\altaffilmark{1}, Daniel Jontof-Hutter, Eric B.\ Ford\altaffilmark{1}}
\affil{Department of Astronomy \& Astrophysics, and Center for Exoplanets and Habitable Worlds, 525 Davey Lab, The Pennsylvania State University, University Park, PA, 16802}
\altaffiltext{1}{NASA Nexus for Exoplanet System Science}



\begin{abstract}
Arnold, Forgan, and Korpela et al. noted that planet-sized artificial structures could be discovered with {\it Kepler} as they transit their host star.  We present a general discussion of transiting megastructures, and enumerate ten potential ways their anomalous silhouettes, orbits, and transmission properties would distinguish them from exoplanets. We also enumerate the {\it natural} sources of such signatures.

Several anomalous objects, such as KIC 12557548 and CoRoT-29, have variability in depth consistent with Arnold's prediction and/or an asymmetric shape consistent with Forgan's model.  Since well motivated physical models have so far provided natural explanations for these signals, the ETI hypothesis is not warranted for these objects, but they still serve as useful examples of how non-standard transit signatures might be identified and interpreted in a SETI context.  Boyajian et al.\ 2015 recently announced KIC 8462852, an object with a bizarre light curve consistent with a ``swarm'' of megastructures. We suggest this is an outstanding SETI target.

We develop the {\it normalized information content} statistic $M$ to quantify the information content in a signal embedded in a discrete series of bounded measurements, such as variable transit depths, and show that it can be used to distinguish among constant sources, interstellar beacons, and naturally stochastic or artificial, information-rich signals.  We apply this formalism to KIC 12557548 and a specific form of beacon suggested by Arnold to illustrate its utility.

\end{abstract}

\keywords{extraterrestrial intelligence --- stars:individual(\object{KIC 12557548}, \object{KIC 8462852}, \object{Kepler-4}, \object{Kepler-5}, \object{CoRoT-29})}

\section{Introduction}

\begin{table*}
\label{tab:anomalies}
\caption{Ten anomalies of transiting megastructures that could distinguish them from planets or stars. }
\begin{tabular}{L{2in}L{2.25in}L{2.25in}}
 Anomaly & Artificial Mechanism & Natural Confounder  \\\hline
Ingress and egress shapes & non-disk aspect of the transiting object or star& exomoons, rings, planetary rotation, gravity and limb darkening, evaporation, limb starspots \\ [6pt]
\hline Phase curves & phase-dependent aspect from non-spherical shape & clouds, global circulation, weather, variable insolation \\ [6pt]
\hline Transit bottom shape & time-variable aspect turing transit, e.g.\ changes in shape or orientation& gravity and limb darkening, stellar pro/oblateness, starspots, exomoons, disks \\ [6pt]
\hline  Variable depths & time-variable aspect turing transit, e.g.\ changes in shape or orientation & evaporation, orbital precession, exomoons \\ [6pt]
\hline  Timings/durations & non-gravitational accelerations, co-orbital objects& planet-planet interactions, orbital precession, exomoons \\ [6pt]
\hline  Inferred stellar density & non-gravitational accelerations, co-orbital objects & orbital eccentricity, rings, blends, starspots, planet-planet interactions, very massive planets \\ [6pt]
\hline  Aperiodicity & Swarms & Very large ring systems, large debris fields, clumpy, warped, or precessing disks \\ [6pt]
\hline  Disappearance & Complete obscuration & clumpy, warped, precessing, or circumbinary disks \\ [6pt]
\hline  Achromatic transits & Artifacts could be geometric absorbers & clouds, small scale heights, blends, limb darkening \\ [6pt]
\hline  Very low mass & Artifacts could be very thin & large debris field, blends \\ [6pt]\hline
\end{tabular}

\end{table*}

Advanced, spacefaring civilizations might have significant effects on their circumstellar environment, including the construction of planet-sized structures or swarms of objects \citep[see][and references therein]{GHAT1}.  Such ``megastructures'' might be detectable both by the starlight they block, and by the midinfrared radiation they emit after reprocessing this light.

The motivations behind the construction of such structures might be obscure, but at least two are general enough to be plausible.  The first is to harvest energy and dispose of it \citep{Dyson60}: the vast majority of the free energy in a stellar system is in the mass of the star itself \citep[e.g.,][]{GHAT2}, and stars naturally provide this free energy via nuclear fusion in the form of starlight.  Large starlight collectors are thus an obvious and long-term means to collect a very large, sustainable energy supply (indeed they may be the {\it only} such means\footnote{The other mechanisms would be to convert the star's mass to energy more efficiently or quickly in some other way, such as feeding it to a black hole, or to have access to ``new physics'' \citep[see][for a discussion.]{GHAT2} }) and large radiators are necessary means to dispose of that much energy after its use.

A second possible motivation, suggested by \citet{Arnold05}, is that large objects could serve as powerful, long-lived, low-maintenance ``beacons'' --- signals of unambiguously intelligent origin that by their very existence delivered a simple ``we are here'' message over long distances in a manner likely to be detected.  These two motivations are not mutually exclusive --- indeed, \citet{Kardashev64} suggested that starlight collection might be motivated by the desire to power radio beacons.

\citet{Dyson60} showed that if a civilization undertook megaengineering projects, the effects on the star would be detectable, and potentially dramatic.  Specifically, he noted that large light-blocking structures around a star would obscure the star, making it dimmer in the optical, and reradiate the collected starlight in the thermal infrared (according to its effective temperature).  This paper focuses on the former effect, but the latter effects would also be observable with modern astronomical techniques \citep{GHAT2,GHAT1}.

\citet{Arnold05} noted that long-term, precise photometric monitoring of stars for transiting exoplanets by {\it Kepler} \citep{Kepler} is effectively a search for alien megastructures while searching for transiting planets, because {\it Kepler} had the capacity not only to detect such structures but the photometric precision to distinguish many classes of megastructures from exoplanets.  In principle, then, an analysis of {\it Kepler} (or similar) data should provide an upper limit to their frequency in the Galaxy.  

Calculating such an upper limit, however, would first require robustly characterizing any and all anomalous signals, of which there are many. Such anomalies are inherently astrophysically interesting, and so deserve careful attention for both conventional astrophysics and SETI.  Here, we describe the signatures that any comprehensive photometric search such structures (perhaps with radial velocity follow-up) should be sensitive to. Describing the details of such a comprehensive search is beyond the scope of this work, but might proceed along lines similar those of the {\it Hunting Exomoons with Kepler} program of \citet{HEK1}.

\section{Distinguishing Features of Megastructures}
\label{sec:features}
\subsection{Signatures of Megastructures}

Below, we discuss six potential aspects of transiting megastrutures that lead to ten observable signatures that would distinguish them from transiting exoplanets.  We enumerate the ten signatures in Table~\ref{tab:anomalies}

\subsubsection{Anomalous Aspects}

\citet{Arnold05} considered the transit light curves of planet-sized artificial objects, such as those that might be used to intercept star light (i.e.\ ``giant solar panels'').  If such objects had non-disk (e.g., triangular) aspects, \citeauthor{Arnold05} argued they could be distinguished from planets by their anomalous transit light curves.  \citeauthor{Arnold05} focused on the opportunities for such an object to serve as a beacon, and noted that transmission power of such a beacon was provided by the star itself, potentially giving such beacons very low marginal transmission costs and very long lives compared to other proposed forms of beacons.  

Arnold described three examples of such beacons. First, he noted that a megastructure with triangular aspect would have anomalous ingress and egress shapes compared to an exoplanet, and that  {\it Kepler} would be able to distinguish the two cases for Jupiter-sized objects.  Second, he considered a series of objects with identical periods whose transits resulted in clearly artificial patterns of spacings and transit depths.  Finally, he considered a screen with louvres which could be rotated to modulate the fraction of stellar flux blocked, producing complex transit light curves between second and third contact that could transmit low bandwidth information, such as a sequence of prime numbers.  

\citet{Korpela15} considered the case of a fleet of structures in a halo around a planet, such as mirrors used to provide lighting to the night side.  If these satellites, as an ensemble, had sufficient optical depth and orbital altitude then they would produce their own transit signature as the planet transited the disk of the star, much like a thick, gray atmosphere.  \citeauthor{Korpela15} showed that {\it Kepler} would not, but {\it James Webb Space Telescope} would, be able to detect the ingress, egress, and transit bottom anomalies from such a satellite swarm around terrestrial planet in the Habitable Zone \citep{Kasting93} of a $V\sim 11$ star.  

A non-spherical megastructure would generate a non-standard light curve in reflection or emission, as well.  For instance, a disk rotating synchronously to keep its surface normal to the incoming starlight (as, perhaps, a stellar energy collector) would present a circular aspect during transit, but a vanishing aspect at quadrature.  Its phase curve, either in reflected or emitted light, would thus have zeros at quadrature, while a spherical object's flux would steadily increase through quadrature toward superior conjunction.  If a transiting megastructure with non-circular aspect and high effective temperature or reflected flux were suspected via ingress and egress anomalies, then a high signal-to-noise ratio (S/N) detection of ingress and egress of its secondary eclipse could confirm its shape and break degeneracies with potentially poorly constrained limb-darkening parameters

\subsubsection{Anomalous Orbits}
\label{astrodensity}

Real exoplanets and stars are appreciably accelerated only by the force of gravity, and so their transit times and durations must obey certain physical constraints parameterized by their impact parameter and the density of the star they orbit \citep{Seager2003}.  \citet{Kipping2014}  described the technique of ``asterodensity profiling'' by which various properties of a star-planet system could be diagnosed via deviations of the stellar density derived via the \citeauthor{Seager2003} relations from the true density of the star (which might be measured independently by other transiting objects or other methods, such as asteroseismology).  \citeauthor{Kipping2014} identified six effects that would create such a discrepancy: orbital eccentricity, blends with other stars, starspots, dynamically generated transit timing variations (TTVs), transit duration variations (TDVs), and very massive planets.\footnote{That is, the usual calculation of a host star's density from transit parameters is made under the assumption that the transiting object has negligible mass, so very massive planets or brown dwarf companions will yield anomalous density estimates.}  \citet{ZuluagaRings} added a seventh: the photo-ring effect.  

Artificial structures, however, might be subject to non-gravitational forces, such as radiation pressure or active thrusts and torques for attitude control and station keeping.  As such, their transit signatures might be distinguished by an ``impossible'' mismatch among the duration and period of the transits, and the stellar density.  To \citeauthor{Kipping2014} and \citeauthor{ZuluagaRings}'s list we therefore here add an eighth asterodensity profiling effect, presumably only applicable to megastructures: significant non-gravitational accelerations (the ``photo-thrust effect'').

In the limit of very small radial thrusts (as in, for instance, the case of radiation pressure on a solar collector), the photo-thrust effect results in an inferred stellar density too low by a fraction $\beta$ equal to ratio of the thrust to the gravitational force from the star that would otherwise keep the structure on a purely Keplerian orbit ($\beta = F_{\rm thrust}/F_{\rm Kep}$, see Appendix~\ref{photothrust}).   In that appendix we also show that planet-sized megastructures with surface densities comparable to common thin metal foils would have photo-thrust effects detectable via asterodensity profiling.

The most extreme case of an anomalous orbit is a static shield, an object held stationary with respect to the star through the balance of thrust (via, for instance, radiation pressure) and gravitational accelerations \citep[a ``statite,''][]{mcinnes1989halo,forward1993statite}.  Such a structure might exist only to collect energy, although the resulting imbalance of outgoing photon momentum would result in a small thrust on the statite-star system (resulting in a ``class A stellar engine'' \citep{Badescu2000} or ``Shkadov thruster'' \citep{Shkadov}), and a warming of the star itself.  Although we currently know of no material with sufficiently low surface density and opacity that could create such a shield around a solar-type star \citep[e.g.][p.32]{kennardthought}, presumably the materials science problem of manufacturing such a substance is easier to solve than the engineering problem of constructing a planet-sized shield.  

In this case there would be no transits to observe, but the shield might obscure a constant fraction of the stellar disk.  \citet{Forgan13} noted that light curves of an exoplanetary transit of a star with such a shield would be anomalously short and asymmetric.  In a reversal of the proposal of Arnold, here it is the apparent aspect of the {\it star} that is non-circular due to alien megastructures, not the transiting object.  As a result, the transit shape anomalies are significantly larger in Forgan's model of an ordinary planet plus a static shield, than in Arnold's model of a non-circular megastructure.

\subsubsection{Swarms}

\label{swarms}
A civilization that built one megastructure might be expected to build more \citep{GHAT1}.  Their host star might therefore be transited by many artificial structures of a variety of periods, sizes, and aspects --- a ``swarm.''  In the limit of a very large number of very small objects, the ensemble might appear as a transluscent screen, and not be easily detected.  Large numbers of larger objects might contribute to a constant, low-level variability that could be mistaken for photospheric noise due to granulation \citep[e.g.,][]{Bastien13} or astereoseismic variations.  Larger objects might generate light curves characterized by aperiodic events of almost arbitrary depth, duration, and complexity.  Such a light curve might require highly contrived natural explanations (although given the number of stars surveyed to date by, for instance, {\it Kepler}, and the rarity of such signals, contrivances might be perfectly warranted).  

\subsubsection{Complete Obscuration}

The most extreme case of a transiting megastructure would be a structure or swarm so large and opaque that it completely occults the star.  In this case there might be a very small amount of scattered light from other components of a swarm, but for the most part the star would go completely dark at optical wavelengths.  In the limit that such a structure or swarm had complete coverage of the star, one has a ``complete Dyson sphere'' \citep[$\alpha =1$ in the AGENT formalism of][]{GHAT2}.  Less complete swarms or structures (as in the case of Badescu and Shkadov's scenarios above) undergoing (perhaps non-Keplerian) orbital motion might lead to a star ``winking out'' as the structure moved between Earth and the star.  In such scenarios, the occulting structure might be detectable at midinfrared wavelengths if all of the intercepted stellar energy is ultimately disposed of as waste heat (that is, in the AGENT formalism, if $\epsilon\approx\alpha$ and $\alpha$ is of order 1).  

\subsubsection{Anomalous Masses}

An artificial structure might have very low mass --- solid structures or swarms of structures could have very large collecting or radiating areas that block significant fractions of starlight, but have no appreciable gravitational influence on their star or planets orbiting it.  Such megastructures would appear anomalous because of the very low densities astronomers would infer from mass measurements via, for instance, radial velocity or TTVs from other transiting objects in the system.  

\subsubsection{Anomalous optical properties}

Most megastructure models invoke of geometric absorbers, and so predict nearly achromatic eclipses.  In contrast, stars are luminous, and brown dwarfs and exoplanets have atmospheres (and, in some cases, dust trails) which can show spectral features in transmission (such as absorption lines and wavelength dependent scattering and absorption).   If a transit signature were gray in both broadband and spectroscopic measurements, this would imply that the object has no detectable region of dust or gas in transmission (respectively).  Note that even in the case of purely geometric absorbers there may be some wavelength dependence in transit depths from limb darkening \citep[and, potentially, diffraction, e.g.,][]{Forgan13}, but this effect is well understood and can be modeled to separate the chromatic effects of the source and transiting object.  

In particular, the exoplanetary interpretation of observations of a large object with very low inferred density (as might be expected for megastructures), would have to invoke a very large atmospheric scale height (of order the radius of the object) and thick atmosphere.  Any such real exoplanet would be an extremely favorable target for transmission spectroscopy, and so the lack of any spectral features or wavelength dependence in such an object's transit signature would be both easily established and extremely difficult to explain naturally.

Observing the object spectroscopically in secondary eclipse would potentially reveal its albedo and composition.  Such measurements can be difficult, but, in combination with the other signatures above, might be sufficient to demonstrate that a megastructure had been detected.

\subsection{Confounding natural sources of megastructure signatures}

Complicating any effort to detect megastrutures is that many of these effects are expected for natural reasons, as well.  For instance, rings or moons produce a non-circular aspect \citep{TusnskiRings,HEK1,ZuluagaRings}.\footnote{Indeed one of the primary outcomes of the precise transit light curves of \citep{Pont2007} was a demonstration that the aspect of HD 189733 {\it b} was indistinguishable from a perfect disk --- a null result, but one which demonstrates the possibility of detection.}  Below, we consider these and other natural origins of the anomalies we describe above.

\subsubsection{Planet-Planet Interactions}

Non-Keplerian motion in real exoplanets can be the result of planet-planet interactions.  Such interactions can generate large TTVs, and even TDVs, especially if the perturbing planet and the transiting planet are in or near a mean motion resonance \citep[MMR,][]{Agol2005,Holman2005} including the 1:1 (Trojan) resonance \citep{FordHolman}.  These interactions also lead to the photo-timing and photo-duration effects in asterodensity profiling (see Section~\ref{astrodensity}).

These interactions can be diagnosed by the fact that they cannot produce arbitrary TTVs --- their patterns are constrained by the families of orbital parameters consistent with long-term stability of the system.

Perturbers near an MMR generally generate a TTV signal that is dominated by a sinusoid, with a period that depends on the proximity of the orbits to an MMR, and an amplitude that depends on the orbital eccentricities and masses of the planets \citep[e.g.,][]{Lithwick2012}.  Diagnosis is especially straightforward if both exoplanets transit and are near an MMR --- in this case each exoplanet perturbs the other, and the two exoplanets exhibit, roughly, opposite TTV signals, with amplitudes that depend on their masses and eccentricities \citep{Kepler-36,Ford11}.  More complex signals can be generated in systems with more than two strongly interacting planets \citep[e.g.,][]{Jontof2014}, but the physical constraints that the system be dynamically stable prevent the natural generation of arbitrary TTVs.  

The maximum observed TTVs and TDVs to date are those of the planets of KOI-142 \citep{KOI-142,Barros2014}, with approximately sinusoidal signals and semiamplitudes of $\sim$12h and $\sim$5m, respectively (the higher frequency components of the TTVs have amplitudes $\sim$ 20m).  Amplitudes significantly in excess of this\footnote{At least, on short timescales.  On longer timescales, TTVs can exhibit much larger amplitudes.} or patterns that deviate strongly from the patterns described above would require highly contrived and possibly unstable configurations of exoplanets.

Another, related effect is variations in transit times due to the displacement of the planet-star system by an outer planet or star, creating a varying light travel time for photons carrying the transit information to Earth \citep{LTTTTV}.  Such an effect is easily distinguished because its signal will be described well by the Keplerian motion of the outer planet, and any perturber massive enough to create noticeable TTVs would be generate a large, potentially observable radial velocity signal on the star.

\subsubsection{Asymmetric Planets}

\citet{Seager2002} and \citet{Carter2010} showed how exoplanetary oblateness (and, so rotation rates and tidal locking) could be probed via the ingress and egress shapes of a transit light curve.  These effects are very small, and to date the only such detection is the marginal one that \citet{Zhu2014} describe.  More extreme deviations from non-circularity, such as that due to a ring system \citep{ArnoldRings,BarnesExoRings,DyudinaRings,OhtaExoRings,ChironRing,TusnskiRings,ZuluagaRings} would be easier to detect, and easier to diagnose (for instance, ring systems should exhibit a high degree of symmetry about some axis, which may not be the orbital axis).

\subsubsection{Nonspherical Stars and Gravity Darkening}

Stars, too, may be non-spherical. Rotation may make them oblate, and a massive nearby companion may make them prolate.   The oblateness effect is usually diagnosed through estimates of the stellar rotation period via line widths \citep{CoRoT-29} or the rotational modulation of the light curve via spots; the prolateness effect requires careful examination of the details of the light curve \citep{Morris2013}.  
 
The dominant effect of a non-disk-like stellar aspect on transit light curves is to potentially generate an anomalous transit duration; the effects on ingress and egress shape are small. Gravity darkening, which makes the lower-gravity portions of the stellar disk dimmer than the other parts, can have a large effect on the transit curves of planets and stars with large spin-orbit misalignment, potentially producing transit light curves with large asymmetries and other in-transit features \citep[first seen in the KOI-13 system,][]{GravityDarkening,Barnes2011}. 

Another, less obvious effect of a non-spherical star is to induce precession in an eccentric orbit, leading to TDVs and transit depth variations (T$\delta$Vs) \citep[such precession can also be caused by general relativistic effects,][]{Miralda02,Pal08}.  

Both gravity darkening and orbital precession can be in play at once, as in the TDVs of the KOI-13 system \citep{Szabo12}.  A more dramatic example seems to be the PTFO 8-8695 system \citep{PTFO8,Barnes13}, which exhibits asymmetric transits of variable depth, variable duration, and variable in-transit shape.  The diagnosis of PTFO 8-8695 was aided by its well known age \citep[$\sim$ 2.65 Myr, aided by its association with the Orion star-orming region,][]{Briceno05}, and its deep transits.  Such dramatic effects would not be expected for older, more slowly rotating objects.

\subsubsection{Starspots and Limb-darkening}

Starspots complicate transit light curves.  When a transiting object occults a starspot, it blocks less light than it would in the absence of the spot, and the system appears to slightly brighten.  Such an effect was seen by \citet{Pont2007} in {\it Hubble Space Telescope} observations of HD 189733 {\it b}. For poorly measured transits, such features can also introduce errors in transit time and duration measurements.  Starspots are also responsible for the ``photo-spot'' effect in asterodensity profiling (see Section~\ref{astrodensity}).

Fortunately, there are several diagnostics for starspots.  One is that the shape of starspot anomalies is a well-known function of wavelength, allowing multi-band measurements to identify them.  Another is that long-baseline precise light curves will reveal the spots' presence as they rotate into and out of view.  If the spots are persistent or appear at common latitudes or longitudes, then repeated transits will reveal their nature. Indeed, such a technique has already allowed for spin-orbit misalignments to be measured for several systems \citep{Deming2011,Nutzman2011,Sanchis2011a,Sanchis2011b}.  

Even for a spot-free star, the ingress and egress shapes of a transit depend on the effects of limb darkening and the impact parameter of the transit.  Misestimations of the host star's properties might lead to inappropriate estimates of limb-darkening parameters which, if held fixed in a fit to a transit light curve, might result in a poor fit to the data, misleading one into believing that the aspect of the transiting object is anomalous.  Limb darkening is a wavelength-dependent effect, so misestimates of it might lead one to incorrectly measure how achromatic a transit depth or shape is.

\subsubsection{Exomoons}

\citet{HEK1} describes many ways in which large moons orbiting exoplanets \citep[or, in the extreme case, binary planets,][]{TusnskiRings,Ochiai14,Lewis15} can leave their signature in the transit timing and duration variations, as the exoplanet ``wobbles'' about its common center of mass with the moon.  If the moon is physically large enough, it can produce its own transit events, creating ingress or egress anomalies, or mid-transit brightenings during mutual events.  The orbital sampling effect can also alter the transit light curve, and produce anomalous TTVs and TDVs \citep{Heller2014}. Such effects can not generate arbitrary TTVs, TDVs, or light curves, and so can be diagnosed via consistency with the physical constraints of the exomoon model.  To date, no such effect has been observed, so any effects of exomoons on light curves is likely to be very small.

\subsubsection{Large Occulters: Ring Systems, Disks, Debris Fields, and Dust Tails}

\label{large}
Large occulters (i.e.\ those with at least one dimension similar to or larger than the size of the star) with extreme departures from circular aspects blur the distinctions among the some of the signatures described above, since there might be no clear delineation between ingress, transit, and egress.  Their transit signatures will be highly complex, and, if their orbital periods are long, might take place over long time frames (months).  Such systems would appear similar in many ways to swarms of artificial objects.

Two major categories of large occulters are ring systems and disks.  A ring system or disk around a secondary object can cause complex or severe dimming, as in the cases of 1SWASP J140747.93-394542.6 \citep{Mamajek12,Kenworthy2015}, an apparent ringed proto-planet around a pre-Main Sequence star,  and EE Cep \citep{EECep2,EECep}, a Be star occulted every 5.6 y by an object with what appears to be a large, almost gray, disk.  A disk that is warped, precessing, or that contains overdense regions can also produce occasional and potentially deep eclipses, as in the cases of UX Ori stars \citep[``Uxors,''][]{Wenzel69,The94,Waters98,Dellemond03} including AA Tau \citep{Bouvier13} and V409 Tau \citep{V409Tau}.  In the case of an inclined, warped, and/or precessing circumbinary disk, the stars' orbits might bring them behind the disk in a complex pattern, as in the case of KH 15 D \citep[an eccentric binary star system occulted by a warped disk,][]{KH15DChiang,KH15DJohnson1,KH15D,KH15DJohnson2}.

Such systems can be very tricky to interpret; indeed each of the three listed above required an ad-hoc model (which in the 1SWASP J1407 case is not even entirely satisfactory).  Adding to the weight of these explanations is that all of these systems are sufficiently young that circumstellar and circumplanetary disks are to be expected.  If such a target were to be found to have a light curve so strange that the best natural explanations were highly implausible, and especially if the target star was clearly too old to host extensive circumstellar or circumplanetary disks, then the ETI hypothesis should be entertained and investigated more rigorously.  

One contrived natural explanation for such a signature might be a debris field: a compact collection of small occulting objects.  Such a field might be a short-lived collection of debris from a planetary collision, or might be collected at the Trojan points of a planet.  Such scenarios might be diagnosed through long-term monitoring of the system, especially after one complete orbital period of the field, or, if the swarm contained sufficiently massive bodies, TTVs or radial velocity measurements \citep{FordGaudi,FordHolman}.

We discuss a fourth category of large occulters, the extensive dust tails of evaporating planets such as that of KIC 12557548 \citep[][]{Rappaport12}, in Section~\ref{KIC1255}.

Finally, disks can occult a portion of the stellar disk, creating an asymmetric transit shape analogous to Forgan's model of a static shield.  Such a disk could be diagnosed via the youth of its host star, thermal emission from the disk appearing as an infrared excess, or direct imaging of the disk with interferometry or coronagraphy.

\subsubsection{Eccentric orbits}

The orbital velocities of planets in eccentric orbits are a function of their orbital phase, so their transverse velocity during transit may be significantly different from that of a hypothetical exoplanet in a circular orbit with the same period.  The ``photo-eccentric effect'' \citep[PE,][]{Photoeccentric1} is the resulting transit duration deviation from that expected from a circular orbit (or, equivalently, a component of asterodensity profiling).  PE has been used to estimate the eccentricity distribution of exoplanets \citep{Moorhead2011}, validate and investigate multitransiting systems \citep{Kipping2012,Fabrycky2014,Morehead2015}, and determine the history of the eccentricities of exoplanets \citep{Photoeccentric3}.

Eccentricity can be diagnosed via secondary eclipse timing, stellar radial velocity variations, and dynamical models to TTVs \citep{Nesvorny2014,Deck2015}. 

\subsubsection{Blends}

Blends occur when one or more stars are either bound or coincidentally aligned on the sky with a star hosting a transiting object. Blends are a major source of false positives for exoplanetary transits, and so might also be for megastructures.

For blends involving planet-sized transiting objects (as opposed to blended eclipsing binary false positives), the primary effect of blending is to dilute the transit signal, which, if unrecognized, to first order leads to an underestimation of the size of the transiting object, but not its shape.  There are second-order effects, however, including the ``photo-blend'' effect \citep[which yields a erroneous stellar density estimate because of an inconsistency in the transiting object radius derived via the transit depth and the ingress/egress durations,][]{Kipping2014} and wavelength-dependent transit depths \citep[if the stars have different effective temperatures, e.g.,][]{Torres04}.  

Blends also confound diagnostics of other natural origins of signs of megastructures by providing a second source of photometric variability, starspots, and spectroscopic signatures such as line widths and chromospheric activity levels.  Identification of many of the signatures mentioned here would be complicated by a blend scenario.  For instance, a stable star with an equal mass background eclipsing binary might appear to show a planet-sized transiting object, but have zero Doppler acceleration (suggesting a sub-planetary mass).

Searches for transiting exoplanets have produced a comprehensive framework both for calculating the blend probability for a given source \citep[e.g.,][]{Morton2011,BLENDER,PASTIS1,PASTIS2}, and identifying individual blends, especially via high resolution imagery, careful examination of spectra, multiband transit depth measurements, and single-band photocenter shifts \citep[e.g.,][]{Torres04,Leger2009,Lillo-Box2014,Desert2015,Everett2015,Gilliland2015}.

\subsubsection{Clouds and small scale heights}

Clouds in exoplanets can be nearly gray, opaque scatterers, and so a high cloud deck can serve to hide the spectroscopic and broadband signatures of the underlying gas in an exoplanetary atmosphere.  A high surface gravity and an atmosphere composed primarily of molecules with large molecular weight will have a very small scale height, and so spectra will be unable to probe the thin atmospheric annulus in transmission.  In either case, an achromatic transmission spectrum would be observed \citep[e.g.,][]{Bean2010,Bean2011,Knutson14}.  A small scale height would not preclude a secondary eclipse spectrum from revealing an object's composition, but clouds might introduce a complication.

Clouds, global circulation, and other weather can also produce time-variable or longitudinally dependent albedos, emissivities, and temperatures, giving planets asymmetric and potentially complex phase curves in reflected and emitted light \citep{Knutson12,Hu15,Kataria14,Koll15}.

\subsubsection{Low density planets}

In practice, objects with unexpectedly small densities have been found.  Densities of $\lesssim 0.15$ g cm$^{-3}$ have been found for multiple {\it Kepler} planets, including Kepler-7{\it b} \citep{Kepler-7}, Kepler-79{\it d} \citep{Jontof2014}, Kepler-51{\it b} \citep{Masuda2014}, Kepler-87{\it c} \citep{Ofir2014}, and Kepler-12{\it b} \citep{Kepler-12}.  These low densities may be related to the heavy insolation they receive from their host stars, and in any case the measured masses are all firmly in the planetary regime.  

If a $R \sim R_{\mbox{\jupiter}}$ megastructure were to be found in a short period orbit around a cool star amenable to precise Doppler work, masses as low as a few Earth masses could be ruled out definitively \citep[the state of the art for planet detections is the possible detection of a 1 Earth-mass planet orbiting $\alpha$ Cen B,][]{AlphaCenBb}.  The implied density would then be $<10^{-3}$ g cm$^{-3}$, and the implied escape velocity would be 8 km s$^{-1}$.  These figures are inconsistent with a gas giant exoplanet: the Jeans escape temperature for hydrogen would be 3000 K, which would be similar to or less than the temperature of a short-period planet.  An alternative natural interpretation could be that the object is a small, terrestrial object with an extended, opaque atmosphere.  

In both cases, the natural interpretation would be amenable to testing via transmission and emission spectroscopy and broadband photometry.

\subsection{Searching for Anomalies, and the Role of SETI}

For concreteness and illustrative purposes, the analyses of \citeauthor{Forgan13}, \citeauthor{Arnold05}, and \citeauthor{Korpela15} assume particular geometries and purposes for their structures.  But one need not commit to any particular purpose or design for such structures --- which, after all, might be beyond our comprehension --- to acknowledge that given enough time and technical ability an old alien civilization might build megastructures orbiting stars, and that these structures might be distinguished from natural objects via the signatures mentioned in Table~\ref{tab:anomalies}. A star might exhibit more than one of the above anomalies, in which case mistaking it for a natural source would be less likely.  A star might, for instance, show many, aperiodic transit signatures of varied shapes (consistent with a swarm of megastructures), varied depths, no wavelength dependence, and no radial velocity evidence of planetary mass objects.   Such an object might completely evade simple natural explanation.\footnote{Of course, such evasions are not necessarily signs of engineering; they are usually a ``failure of imagination'' \citep{Clarke3}.  For instance, the ``impossible'' transiting multiple system KIC 2856930 \citep{KIC2856} has eclipses that have so far defied many attempts at physical explanation of increasing contrivance, up to and including quadruple star scenarios with unlikely period commensurabilities.  The invocation of megastructures does not, however, appear to provide any explanatory power to the problem, and so the likeliest solution remains a hitherto unconsidered natural complication.}

Such an object would also escape easy notice because the brightness variations would not necessarily fit a standard transit profile (lacking periodicity, expected durations and shapes, etc.)  It is possible that only a free-form search for ``anything unusual'' across an entire photometric data set would identify such objects, such as by a non-parametric or nonlinear, automated search \citep[e.g.,][]{EBAI,Richards2011,Walkowicz2014} or a human-eyeball-based, star-by-star effort \citep{PlanetHunters1}.  

Given the number of stars observed by {\it Kepler}, and, soon, {\it TESS}, LSST, and other efforts, many unusual transit signatures will undoubtedly be found.  For instance, CoRoT-29{\it b} shows an unexplained, persistent, asymmetric transit --- the amount of oblateness and gravity darkening required to explain the asymmetry appears to be inconsistent with the measured rotational velocity of the star \citep{CoRoT-29}.  \citeauthor{CoRoT-29} explore each of the natural confounders in Table~\ref{tab:anomalies} for such an anomaly, and find that none of them is satisfactory.  Except for the radial velocity measurements of this system, which are consistent with CoRoT-29{\it b} having planetary mass, CoRoT-29{\it b} would be a fascinating candidate for an alien megastructure. 

Until all such unusual objects are identified and explained naturally in a given survey, no upper limit on alien megastructures can be robustly calculated.  Most such signals will, presumably, be natural, and represent unexpected or extremely unlikely phenomena --- alien megastructures should be an explanation of last resort.  But even while natural explanations for individual systems are being explored, all of the objects displaying the most anomalous signatures of artificiality above should be targets of SETI efforts, including communications SETI and artifact SETI \citep{GHAT1}.  

\section{KIC 12557548,  and Other Evaporating Planets as Illustrative Examples of Potential SETI Targets} 
\label{sec:evaporation}
\subsection{Discovery and Evaporating Planet Model}
\label{KIC1255}

\begin{figure}
\plotone{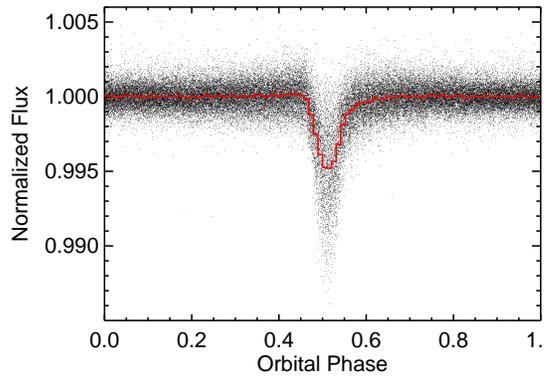}
\caption{Phase-folded photometry of KIC 1255 from all long cadence {\it Kepler} data.  The solid red curve represents the mean flux in each of 96 evenly spaced flux bins.  We have normalized the photometry for each orbit by the mean PDCSAP flux level outside of transit$(|\mbox{phase}-0.5|>0.1)$. After Figures 3a and 3b from \citet{Rappaport12}. \label{curve}}
\end{figure}

\citet{Rappaport12} announced KIC 12557548$b$ (\kic\ for short), an apparently evaporating planet with a 16 hr period discovered with the {\it Kepler} observatory.  Consistent with Arnold's prediction, its transit depths vary, even between consecutive transits ``from a maximum of ∼1.3\% of the stellar flux to a minimum of ∼0.2\% or less without a discernible rhyme or reason'' \citep{Croll14}.  Further, Figure~\ref{curve} shows how the transit light curves ``exhibit[] an obvious ingress/egress asymmetry, with a sharp ingress followed by a longer, more gradual egress'' \citep{Croll14} consistent with Forgan's model.  

In the model of \citeauthor{Rappaport12}, \kic\ is a small planet, similar in size to Mercury, disintegrating under the intense insolation of its parent star.  Ablated material forms an optically thick cometary coma and tail that with a decidedly non-circular aspect, creating the asymmetric transit light curve.  The disintegration is stochastic, with a characteristic timescale longer than one transit (since the transit shape appears consistent) and shorter than one orbit (since the transit depths can vary by a factor of $>6$ between transits).  

\citet{Brogi12} found that the details of the transit shape seem to be well fit by a model invoking a variable, cometary dust cloud.  This includes an apparent brief period of brightening prior to ingress, which they explain as the result of forward scattering of starlight by dust.  \citet{Budaj12} comes to similar conclusions, and also finds evidence for forward-scattered brightening after egress and put constraints on the dust particle size of 0.1--1 $\micron$, with variation along the cometary tail.  \citet{Perez13} find that if the disintegrating planet model is correct, \kic\ is likely in its final stages of its existence, having already lost $\sim 70\%$ of its mass and now being little more than an iron-nickel core.  \citet{Kawahara13} find a small signal in the transit depths at the period of the rotation period of the host star, which they interpret as evidence that the evaporation is correlated with stellar magnetic activity, but which \citet{croll2015} attribute to starspots.

\citet{vanWerkhoven14} performed a detailed analysis of the transit depth time series (DTS), and found only two significant departures from randomness:  long ``quiescent'' periods of $\sim 30$ transits with depths $<0.1\%$, and a few ``on-off'' sequences of alternately deep and shallow transits.  They find that a 2D, two-component model is required to explain the detailed transit shape.

\subsection{KIC 12557548 as Mega-engineering}

The success of the evaporating planet model means that \kic\ has no need of the ETI hypothesis at this time, but we find it to be a useful illustration for a discussion of how similar systems without good natural explanations might be modeled.

The particular light curve shape for \kic\ is not a perfect fit for either the \citet{Arnold05} or \citet{Forgan13} models, but a combination of them nearly works.  The brightening of the system shortly before or after transit can be explained as a glint or forward scattering of starlight from the megastructure.  The asymmetric profile could be a result of either a static occulter, as in Forgan's model, or a highly non-circular aspect, as in Arnold's model.   In particular, a very long triangular aspect structure that enters transit with the wide base first might produce the asymmetric signature seen.     

The variable depths could be explained by active control of the structure to convey low bandwidth information, as in Arnold's model, or with a ``fluttering'' structure that has lost attitude control and is tumbling, folding, or spinning in a chaotic manner, and thus presenting a highly variable aspect to the Earth. 
    
There are difficulties with this model: a large, non-spherical object with such an aspect and short period should be subject to significant tidal torques, and potentially internal dissipation of energy as it folds, and might be expected to quickly achieve spin-orbit synchronicity with its host star.  We note, however, that such a structure might be subject to significant non-gravitational forces, such as magnetic field interactions, radiation pressure or active control, and so never stabilize its orientation.  Or, like a flag flapping in the breeze, a structure might enter periods of semi-periodicity or quiescence while still having an overall chaotic nature with some characteristic timescales.  

Of course, such a structure might also be expected to produce variable transit light curve shapes in accordance with its variable aspect.  We note that due to the shallow depths of its transits, the light curve shape of \kic\ that has been modeled to date is that of an {\it average} transit, and that individual transits may actually exhibit significant variation from this shape.  In addition, there are ways to maintain an aspect while varying the cross section of a structure (e.g.\ a flat, tall triangle spinning about and orbiting in the direction of its long axis) .

\subsection{Other Examples of KIC 12557548-like Phenomena}

\citet{Rappaport13} announced that KOI-2700 showed very similar behavior to \kic, including a very similar, distinctly asymmetric light curve.  In this case, the transit extends to $\sim 25\%$ of the orbit, and the depth variations are secular, weakening by a factor of 2 over the course of the {\it Kepler} mission.  They note that this discovery shows ``that such objects may be more common and less exotic than originally thought.''  Unfortunately, the small transit depths of KOI-2700 prohibit the detailed analysis afforded by \kic.  They note that only very low mass planets $(M\lesssim 0.03 M_\oplus)$ should produce cometary tails of this sort detectable by {\it Kepler} (due to their low surface gravity), so a useful density measurement is not possible.

Most recently, \citet{EPIC2016} announced the discovery of a similar phenomenon in K2-22$b$ from the {\it K2} mission, having a 9.15 hr orbital period and highly variable transit depths.  In addition to the cometary dust tail, their model also includes a leading trail, accounting for a long pre-ingress tail to the transit light curve.  One of their observations also indicates a tentative, shallow wavelength-dependence in transit depths, which can be naturally explained by dust scattering.

\subsection{Diagnosing KIC 12557548-like Objects as Artificial}

The most definitive of our diagnostics for engineering, a density measurement, is impractical for \kic\, , KOI-2700, and K2-22$b$, because the masses expected from the planetary models are too low to produce detectable radial velocity signatures given how faint and active the host stars are.  

Fortunately for SETI efforts, the wavelength dependence of the depths of anomalous transit signals is a matter of considerable astrophysical interest for purely natural reasons, since disks and cometary tails are typically composed of dust and gas, which should exhibit a wavelength dependence diagnostic of the dust grain size and gas composition.

Indeed, \kic\ has been subject to considerable effort along these lines.  \citet{Croll14} observed the target in K$^\prime$ band ($\sim 2.15\micron$) at CFHT simultaneously with {\it Kepler} (in the broadband optical near 0.6 $\micron$) and found no wavelength difference (the depth ratio between the two bands was 1.02$\pm 0.2$)\footnote{\citet{Croll14} also obtained simultaneous {\it Hubble Space Telescope} and {\it Kepler} observations of KIC 12557548, but these occurred during ``quiescence'' and the transits were not detected with either instrument.}  This would imply that the grains in the tail must have a characteristic size $>0.5 \micron$, which may be consistent with the forward-scattering explanation for the pre-transit brightening offered by \citet{Brogi12} and \citet{Budaj12}.  Schlawin et al.\ (2015, submitted; private communication) has also observed \kic\ with IRTF SpeX in the NIR and MORIS in the r$^\prime$ band ($\sim$0.6 $\micron$) simultaneously over 8 nights in 2013 and 2014, and their differential spectroscopy also shows a flat spectrum. 

Interestingly, recent results announced by Haswell\footnote{``Near-ultraviolet Absorption, Chromospheric Activity, and Star-Planet Interactions in the WASP-12 system,'' Haswell et al., Exoclimes {\sc iii}: The Diversity of Exoplanet Atmospheres, 201 November, citing ``Bochinski \& Haswell, et al.\ (2014 in preparation).''} suggest that a large (factor of $\sim 2$) wavelength dependence in transit depth for \kic\ may be seen in the optical from $g$ to $z$ bands, consistent with an ISM extinction law.  It is unclear if these and future results will clarify previous work or throw it into doubt. 

If the transit depths ultimately prove to be achromatic from the infrared through the optical, including an absence of the line absorption that should accompany the gas and evaporated dust, then the ETI hypothesis for this object may need to be reconsidered.

\section{\wtf\ as a SETI Target}

\subsection{Discovery and Initial Characterization of \wtf}

\citet{TabbyWTF} recently announced the discovery of an extraordinary target in the {\it Kepler} field, \wtf\ (KIC 8462, for short).  We briefly summarize their findings below.

Over three years ago Planet Hunter\footnote{Planet Hunters is a citizen science project, in collaboration with Zooniverse, to classify \kepler\ light curves \url{http://www.planethunters.org}} volunteers noticed KIC 8462 as having peculiar variations in its observed flux, with losses over 20\%.  Figure~\ref{fig:8462} shows the full light curve and various zoom levels of such events. These events are extraordinary and unlike any other stellar transit (occultation) events in the {\it Kepler} dataset.

\wtf\ appears to be an F star in the {\it Kepler} field. Its optical spectrum is typical of a Main Sequence (or very slightly evolved) star. Its brightness is consistent with a distance of $\sim$600 pc, and it shows low-level, quasi-periodic variability of 0.5 millimag with a period of $\sim$ 0.88 day, likely due to rotation (its rotational broadening is consistent with a $\sim 1$ day rotation period).
 
\citeauthor{TabbyWTF} confirmed that the ``dipping'' events of KIC 8462 are {\it not} due to instrumental/reduction artifacts in the data: the observations shown in Figure~\ref{fig:8462} are certainly astrophysical in origin. In an attempt to quantify the uniqueness of \wtf's light curve, \citeauthor{TabbyWTF} performed a search through the entire \kepler\ data set of $\sim$100,000 stars to identify similar objects.  The search identified over 1000 objects with $>10$\% drop in flux lasting at least 1.5 hr (with no periodicity requirement). Visual inspection of the resulting light curves revealed that the sample comprises only eclipsing binaries, heavily spotted stars, and KIC 8462852.

\subsection{Difficulties Explaining \wtf\ Naturally}

\begin{figure*}[ht]
\includegraphics[width=3in]{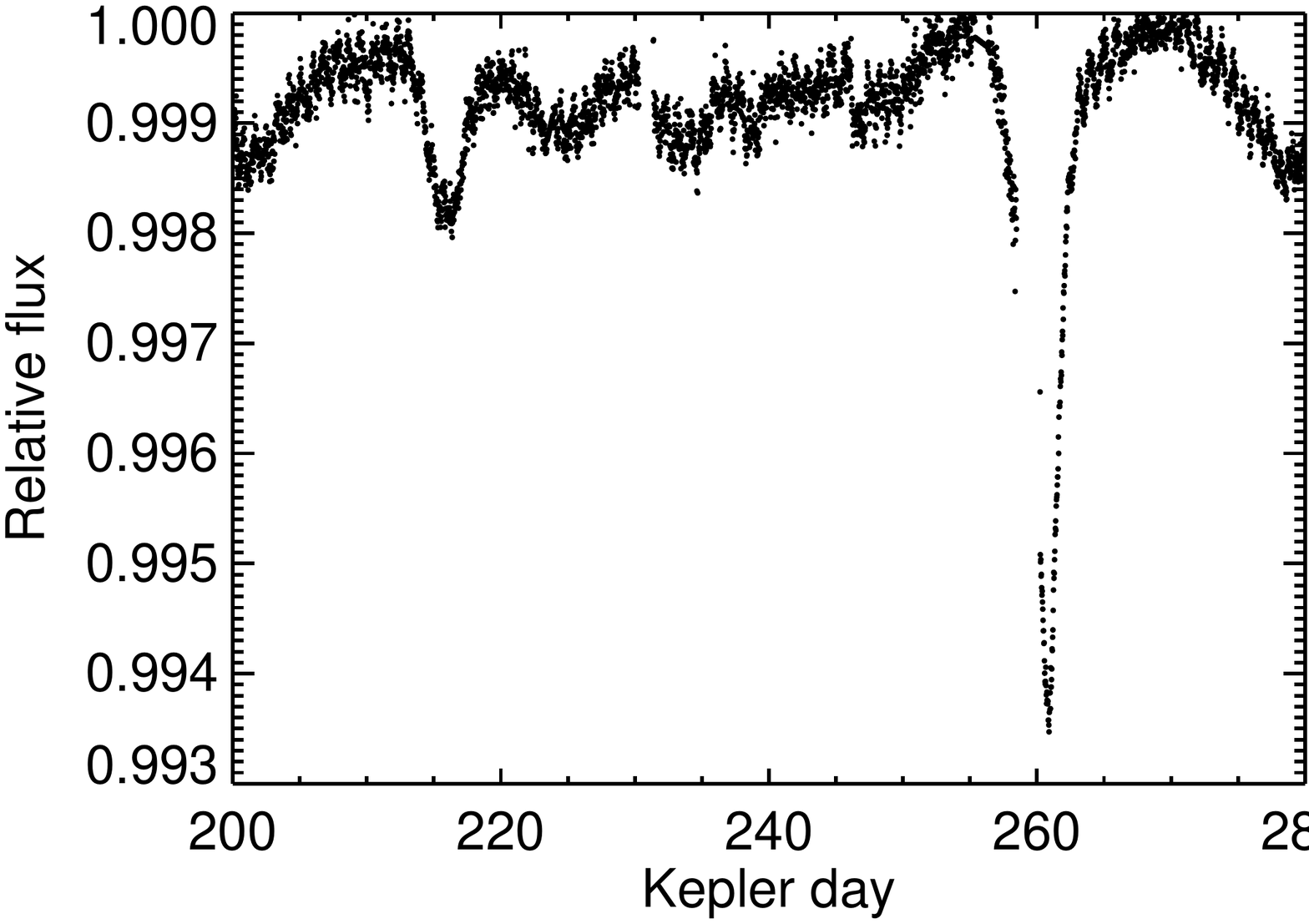}
\includegraphics[width=3in]{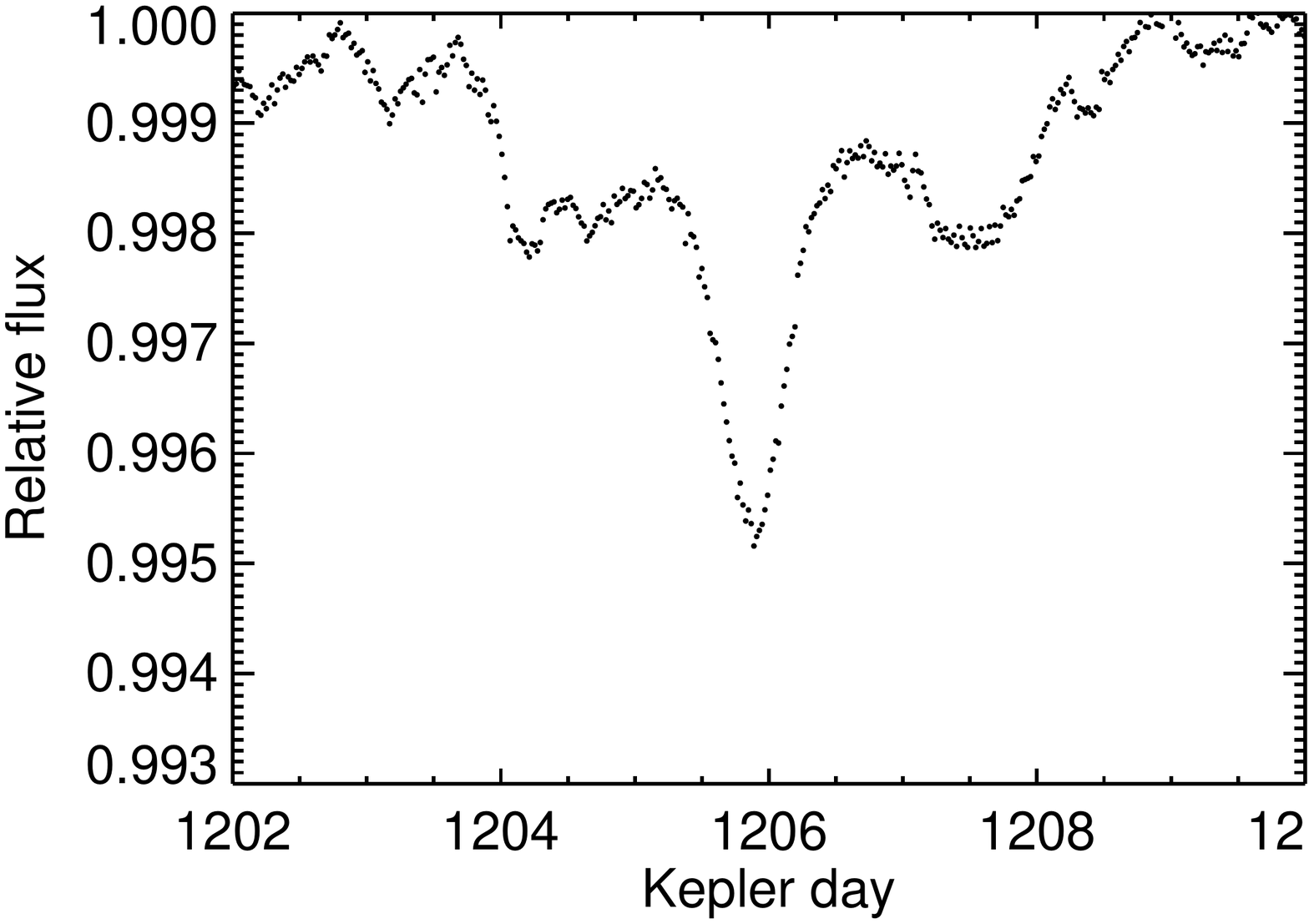}\\
\includegraphics[width=3in]{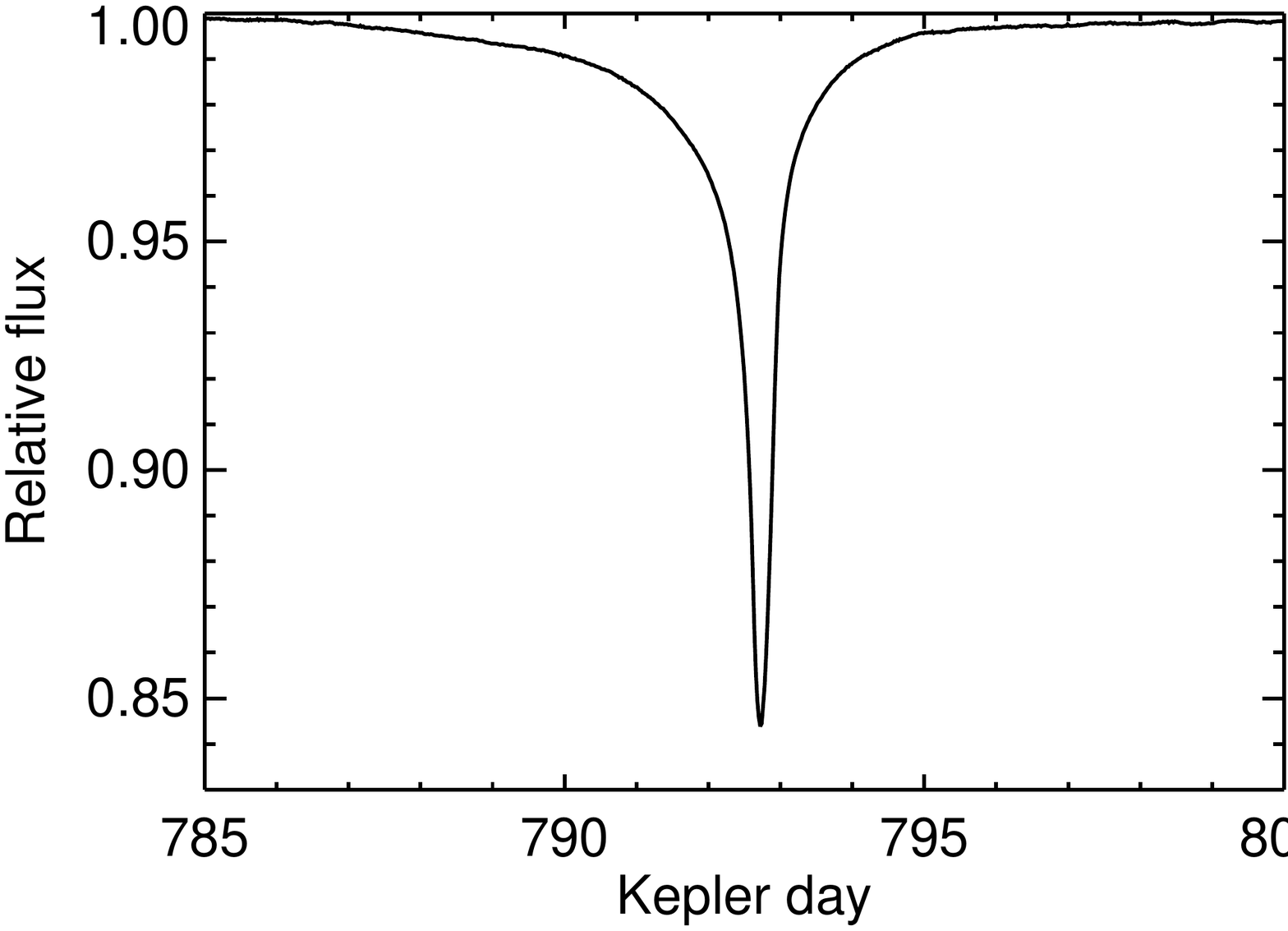}
\includegraphics[width=3in]{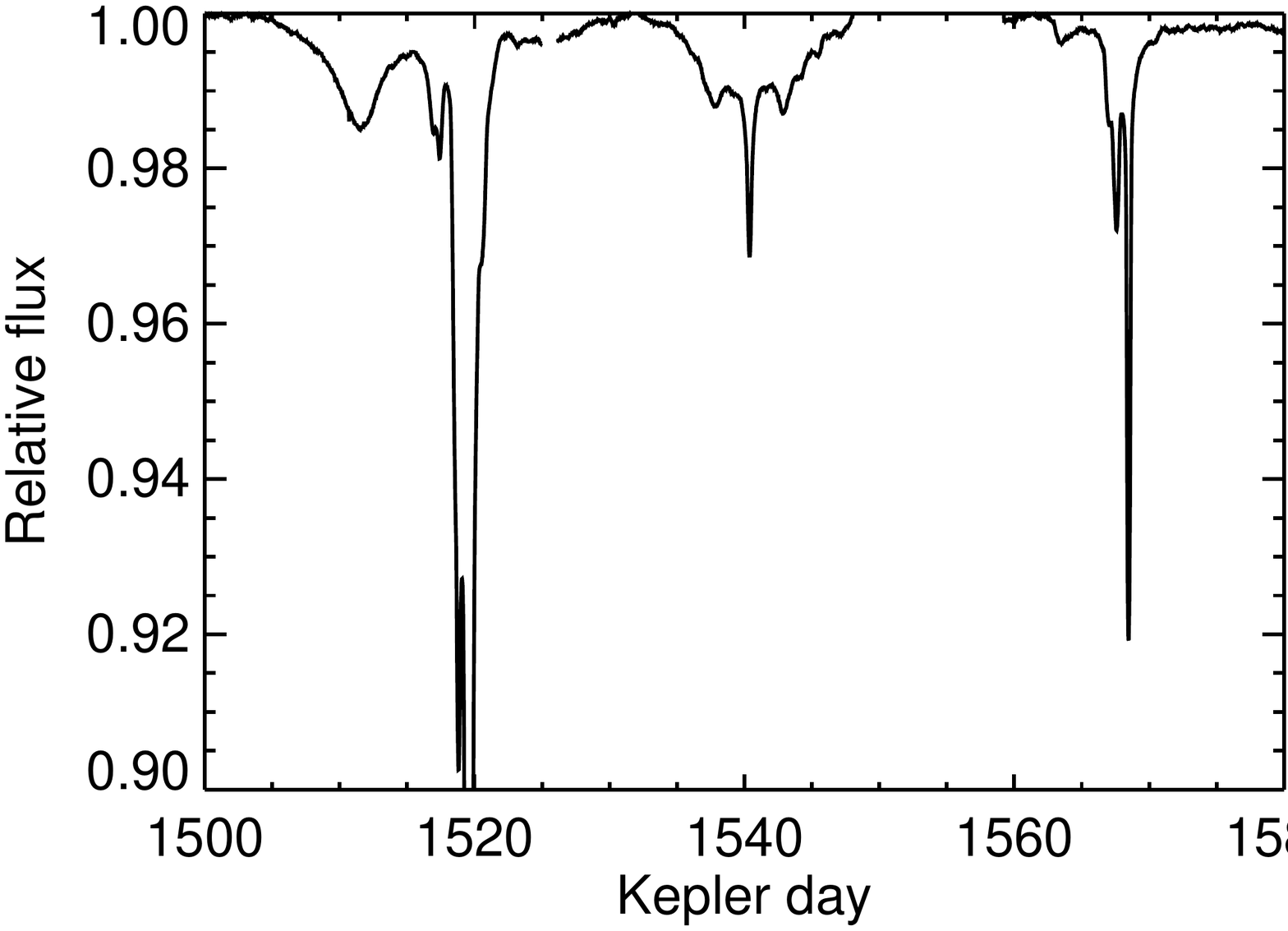}
\caption{\label{fig:8462}
Four details of the \wtf\ light curve.  {Top left}: the high-frequency  ``noise'' is likely due either to rotationally modulated surface inhomogeneities, but the two deeper events at days 216 and 262 are due to something else.  There is also additional variation at the 0.05\% level that persists throughout the light curve, which may be due to typical F star granulation.  {Top right}: a shorter, complex event. {Bottom left}: a deep, isolated, asymmetric event in the {\it Kepler} data for KIC 8462852.  The deepest portion of the event is a couple of days long, but the long ``tails'' extend for over 10 days.  {Bottom right}: a complex series of events.  The deepest event extends below 0.8, off the bottom of the figure. It is unclear if the event at day 1540 might be related to the event at day 1206 from the upper right, which is almost 10 times shallower but has similar shape and duration.  This shape is not repeated elsewhere in the light curve.  Note the differences in scales among the panels.  After Figure 1 of \citet{TabbyWTF}.} 
\end{figure*}

\citeauthor{TabbyWTF} struggled to find a natural explanation for KIC 8462, noting that most of their suggested scenarios ``have problems explaining the data in hand.''   To explain the events as transits, one must apparently invoke a large number of individual transiting objects.  The durations of the events and the lack of repetition require the objects to be on long-period orbits.  A depth of 22\% for the deepest event implies a size of around half the stellar radius (or larger if, like a ring system, the occulter is not completely opaque).  The asymmetries imply that either star or the occulter deviate significantly from spherical symmetry.  The extraordinary event at (BJD-2454833)=793 in Figure~\ref{fig:8462} typifies all three of these qualities.

The complexity of the light curves provide additional constraints:  for a star with a uniformly illuminated disk and an optically thick occulter with constant shape, the shape of the occulter determines the magnitude of the slope during ingress or egress, but not its sign: a positive slope can only be accomplished by material during third and fourth contact, or by material changing direction multiple times mid-transit (as, for instance, a moon might).  The light curves of KIC 8462 clearly show multiple reversals (see the events between (BJD-2454833)= 1500 and 1508 in Figure~\ref{fig:8462}), indicating some material is undergoing egress prior to other material experiencing ingress during a single ``event.''  This implies either occulters with star-sized gaps, multiple, overlapping transit events, or complex non-Keplerian motion.  

The large number of events requires there to be a large number of these occulters --- at least 8 just from the events shown in Figure~\ref{fig:8462}, plus an uncertain number from lower-level events (but at least another 8).  

Explanations involving large ring systems are appealing (see Section~\ref{large}), but the deepest events of KIC 8462 are separated by years (with no periodicity).  Also unlike, for instance, 1SWASP J1407, the KIC 8462 events do not occur symmetrically in time as one would expect from a giant ring system as the leading then trailing parts of the ring occult the star.  In addition, explanations invoking rings and disks would seem to be excluded by the star's lack of IR excess, lack of emission consistent with accretion, and large kinematic age \citep{TabbyWTF}.    

\citeauthor{TabbyWTF} suggest several explanations, but settle on a ``family of exocomet fragments, all of which are associated with a single previous breakup event'' as the one ``most consistent with the data.''

\subsection{An Extraordinary Hypothesis for an Extraordinary Object}

We have in KIC 8462 a system with all of the hallmarks of a Dyson swarm (Section~\ref{swarms}): aperiodic events of almost arbitrary depth, duration, and complexity.  Historically, targeted SETI has followed a reasonable strategy of spending its most intense efforts on the most promising targets.  Given this object's qualitative uniqueness, given that even contrived natural explanations appear inadequate, and given predictions that {\it Kepler} would be able to detect large alien megastructures via anomalies like these, we feel is the most promising stellar SETI target discovered to date.  We suggest that KIC 8462 warrants significant interest from SETI in addition to traditional astrophysical study, and that searches for similar, less obvious objects in the {\it Kepler} data set are a compelling exercise.  

Of course, there may have been many more \wtf-like objects imaged in the {\it Kepler} focal plane that {\it Kepler} failed to discover, because they were not chosen to be among the $\sim$ 100,000 targets to have photometry downloaded to Earth. Likewise, many rare and unexpected targets such as KIC 8462 will also be present in the fields of view of {\it WFIRST}, {\it TESS}, and PLATO, which adds further weight to arguments emphasizing the importance of downloading all data from these future missions, rather than only postage stamps around prime targets.

\section{Distinguishing Beacons From Natural Signals Via Their Information Content} 

\label{sec:beacons}

\subsection{The Normalized Information Content, {\it M}, of Beacons and More Complex Signals}

Two categories of signals from ETIs that we might expect to detect are ``beacons'' and ``leaked'' communication.  The former might be employed by ETIs seeking to be discovered by other intelligent species, and so might be obvious, easily detected, simple, and unambiguously artificial.  These qualities make beacons the focus of many SETI efforts \citep[e.g.][and many efforts since then]{SETI,WaterHole}.  Indeed, pulsars appeared to exhibit many of these qualities, and until its physical nature was deduced the first pulsar discovered was jocularly referred to as ``LGM-1'' (for ``Little Green Men'') by its discoverers \citep{Hewish68,LGM-1}.

By contrast, leaked communication, since it is not intended to be discovered or interpreted by humans, might have none of these qualities.  In particular, it might be characterized by high bandwidth and/or high levels of compression, making its signal highly complex with an extremely high information content.  For a signaling process of a given bandwidth, increasing the information content results in the measurements more closely resembling a random signal, potentially thwarting attempts to distinguish an artificial signal from the natural variability of an astrophysical source.  

If an alien signal is detected, it will be important to determine if it is a beacon, whose purpose and message might be discernable, or a much more complex signal, which might be beyond our comprehension.\footnote{Given that ETIs might be arbitrarily more technologically and mathematically more advanced than us, interpreting a complex signal might be an impossible task, akin to Thomas Edison attempting to tap the telecommunication signal carried by a modern optical fiber cable.  Even if we were to somehow notice, intercept, and successfully record the signal, there is no guarantee we would be able to decipher it.}  A first step, then would be to characterize the complexity of the signal.  Similarly, the case for a potentially alien signal being an artificial beacon would be strengthened if its information content were low but non-zero, and not maximal (as in the case of pure noise).  

SETI therefore would benefit from a quantification of signal complexity that clearly distinguishes beacons, signals with zero information content, and signals with maximal information content.  

A signal can serve as both a beacon and a high-information-content signal by being simple in the time domain but complex in the frequency domain, or vice versa.  For instance, a simple sinusoidal signal could act as a carrier wave, and small variations in the amplitude and/or frequency of the wave could carry complex information.  An ideal statistic of information content should therefore be applicable in both the time and Fourier domains, and be able to give different values in each.

We have chosen to use the Kullback-Leibler divergence $K$ as the basis of our metric, and we describe its calculation from our discrete DTS in detail in Appendix~\ref{sec:kl}.  Relevant here is that it computes the {\it relative} entropy between two {\it distributions}.  In the time domain, we use the probability density function (PDF) of the measured signal, produced from the DTS via kernel density estimation (KDE), and compare to synthetic PDFs of constant and uniformly random signals.  In this case, $K$ has a small value for constant signals ($\delta$-function distributions) and large values for uniformly random signals (uniform distributions).  In the frequency domain, we use the discrete Fourier transform in place of the PDF.  In this space, $K$ has a small value for signals with power at a single frequency (or constant signals) and maximal values for white noise.  This formalism can also be expressed in other bases in which information might be transmitted.

To help interpret the $K$ values we compute for given time discrete series, we propose in Appendix~\ref{sec:m} the {\it normalized information content}, which quantifies the complexity of a signal in the time or Fourier domains on a simple scale from zero (no information) to one (maximal information), with beacons having intermediate values.  The value of $M$ measured for a given signaling process will depend on many factors, including the precision of the measurements and the length of portion of the signal observed.  Measuring a low value of $M$ means that the signal {\it appears} constant at a given precision, and measuring a very high value means that it {\it appears} to be uniformly random.

\subsection{Time Series Analysis of Beacons and Real Transiting Systems}

To illustrate the normalized information content $M$, we apply it to several different cases, enumerated below.  The source code for these calculations, written in R, is available as supplemental electronic tar.gz files associated with this paper.

We use the {\it Kepler} time series of the apparently evaporating planet \kic\ to illustrate a complex, near maximal signal, as might be expected from a stochastic natural source or an information-rich signal transmitted via an Arnold beacon.  We use the {\it Kepler} time series for Kepler-4{\it b} \citep{Kepler-4} as an ``ordinary'' transiting planet  (so, having near-zero information content) because it has a S/N very similar to \kic\ and so makes a good comparison.    We also consider the specific beacon signal proposed by \citet{Arnold05} to illustrate its intermediate relative $M$ values (at least, in the high-S/N case.)

To illustrate the effects of S/N on the detectability of beacons and entropy measurements, generally, we also consider Kepler-5{\it b} \citep{Kepler-5}, which has a much deeper transit and so is measured at much higher S/N than Kepler-4{\it b}. We also consider a hypothetical version of \kic\ observed at a similarly high S/N but with the same measured depth values, and the same Arnold beacon as in the lower S/N case.  

Although our depth measurements are slightly heteroskedastic, our derived uncertainties in transit depth of real systems are sufficiently close to constant that in what follows we choose to use the mean of the uncertainties for a given system as characteristic of the noise.

\subsubsection{\kfour\ and \kfive}

\label{sec:kfour}

\begin{figure*}[t]
\begin{center}
\plotone{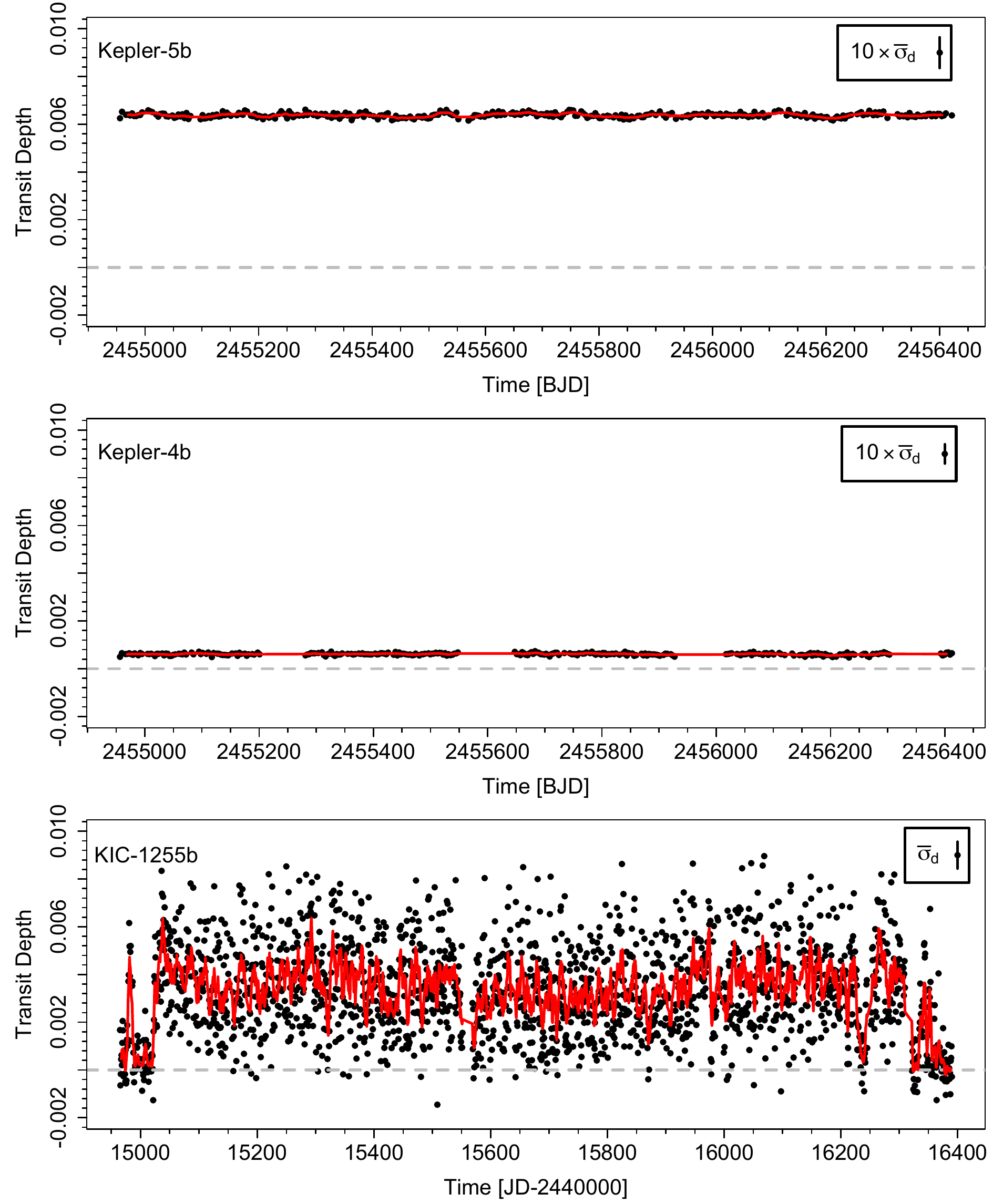}
\caption[DTS of \kfour\ and \kfive]{Depth time series of \kfive\, (black points, top), \kfour\ (middle), and \kic\ (bottom), shown at the same vertical scales (the horizontal scales are slightly different).  We show a moving average (width = 7 transits) in red.  \kfour\ is missing data from quarters 8, 12, 16, and part of quarter 4 due to instrument failure on the spacecraft.  \kic\ lacks data from quarters 0 and 17.  Characteristic uncertainties are indicated in the legends (note the inflation factors applied for clarity).}
\label{fig:k4}
\end{center}
\end{figure*}

\begin{figure*}[t]
\begin{center}
\plottwo{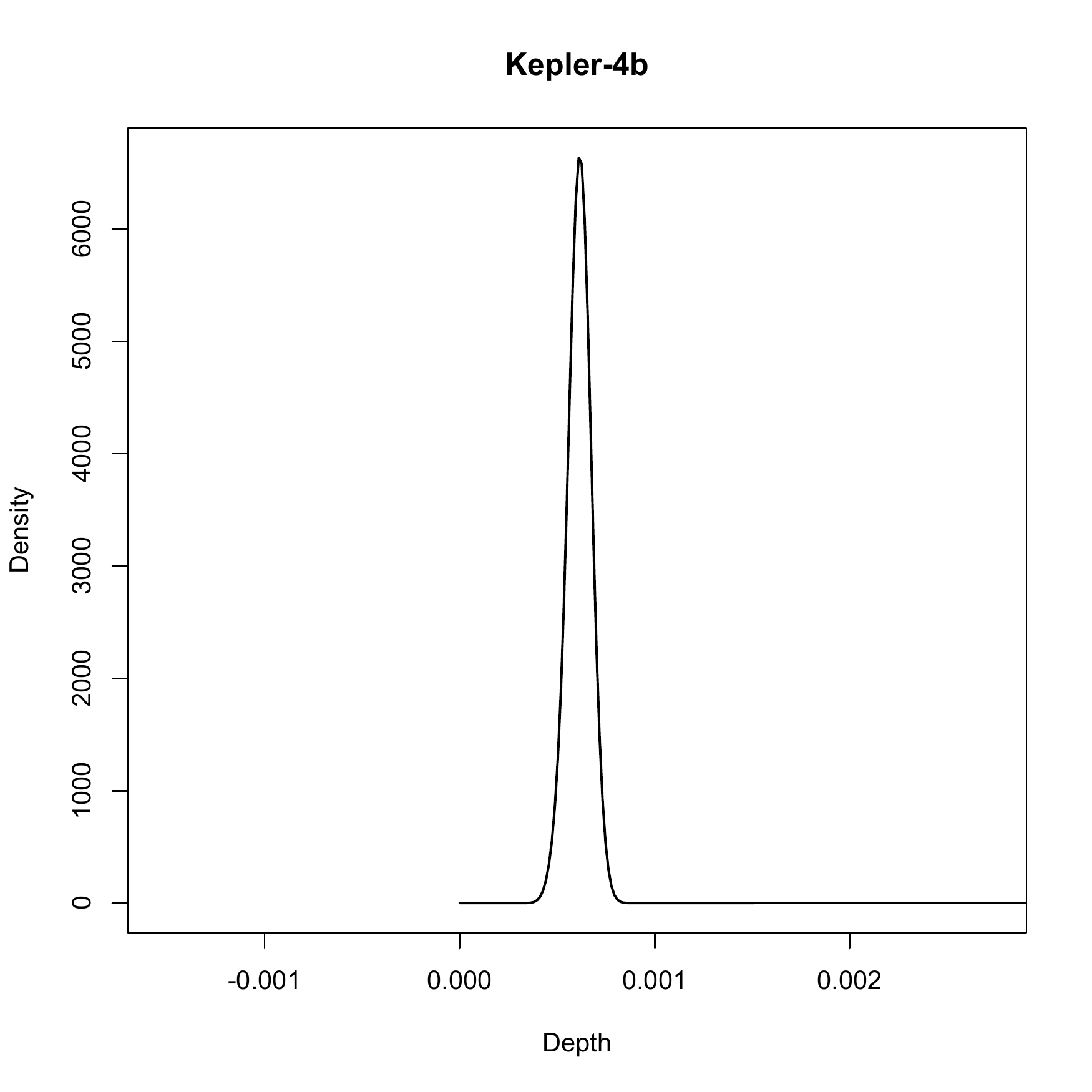}{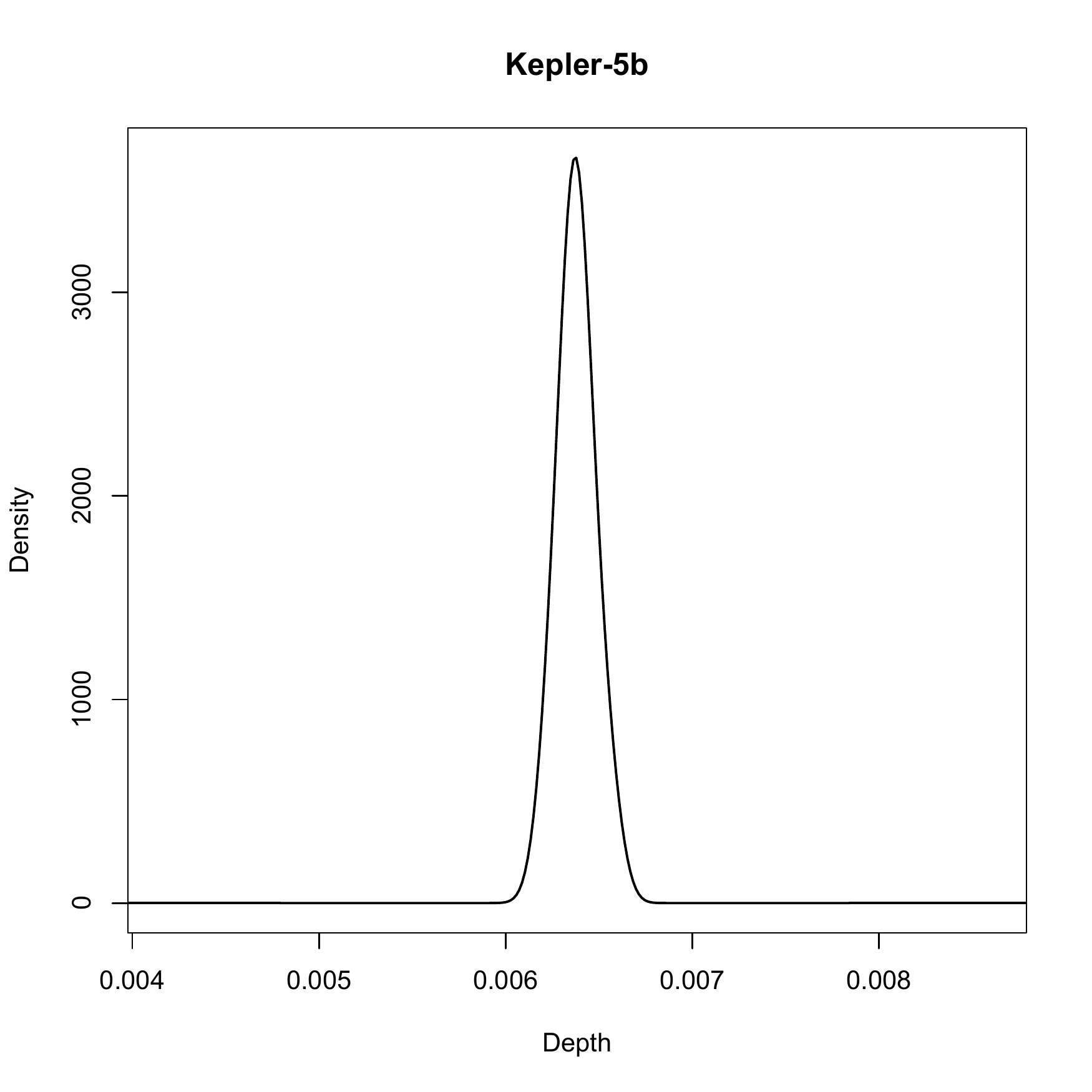}
\caption[PDF of \kfour\ and \kfive]{{Left:} PDF of the depth time series of \kfour. Width of convolving kernel is the mean measurement uncertainty of \kfour. {Right:} PDF of the DTS of \kfive. Width of convolving kernel is the mean measurement uncertainty of \kfive}
\label{fig:k5}
\end{center}
\end{figure*}

In order to illustrate an ``information free" DTS, we chose \kfour, a $\sim4R_\oplus$ planet with a 3.21 day period orbiting a 1.2 $M_\odot$ star, and \kfive, a $\sim 1.4 R_{\mbox{\jupiter}}$ planet with a 3.5 day period orbiting a 1.4 $M_\odot$ star. Their transit depths are $\mu_{d} = 728 \pm 30$ ppm and $6600 \pm 60$ ppm, respectively. Figure \ref{fig:k4} (middle) shows the DTS for \kfour, for quarters 1--17 (excepting 8, 12, and 16). From this we see that the \kfour\, transits are very regular, and so we anticipate little information content in the time series. The gaps in the \kfour\, DTS are due to missing long cadence data from quarters 8, 12, and 16, part of quarter 4, and regular instrument shutdown times. The mean transit depth of \kfour\, is a bit lower than the mean transit depth of \kic, our benchmark ``false positive" case, but since the star is brighter it has a comparable S/N to \kic\, which makes it an ideal comparison target.

We generated the \kfour\ and \kfive\ DTS's from all available quarters \kepler\ long-cadence data for these targets. We downloaded the light-curves from the Mikulski Archive for Space Telescopes (MAST) and removed low-frequency  variability using the Pyke function {\tt kepflatten} \citep{pyke}. We then ran the flattened light curves through the {\tt autoKep} function of the Transit Analysis Package \citep{tap} to identify the planetary transits, which we then folded and jointly fitted to a transit light curve model using {\tt exofast} \citep{exofast} and the stellar parameters found on the \kepler\, Community Followup Observing Program (CFOP).\footnote{\url{https://cfop.ipac.caltech.edu/home/}}  Having solved for the parameters of the system, we then re-fit each transit individually fixing all transit parameters (using very narrow priors) except transit depth in the {\tt exofast} fitting.  In a few cases, we identified anomalous fits (reduced $\chi^2 >5$), which we rejected.

In principle, hypothetical unseen planets in the \kfour\ or \kfive\ systems could affect the fitting of the transit DTS, since we have forced the transit centers to fit a strictly linear ephemeris. However, we are not motivated to perform more detailed investigations considering the large parameter space for undetected planets and the results of a Durbin-Watson test\footnote{This test finds $p$ values for the alternative hypotheses that the true autocorrelation in the DTS is greater than and less than zero.  For \kfour\ we find a Durbin-Watson statistic of 1.8653, so $p=0.1036$ and 0.8964 for positive and negative autocorrelations, respectively. For \kfive\ we find a statistic of 1.9609, so $p=0.3328$ and 0.6672, respectively. } that show no evidence for a significant positive or negative autocorrelation in either dataset.

This method generated a DTS for \kfour\ and \kfive, with 282 and 351 transits, respectively, including depths, depth uncertainties, and transit center times. The DTS's and their PDF's of the \kfour\ and \kfive\ light curves are shown in Figures \ref{fig:k4}--\ref{fig:k5}.  Our DTSs are included in our supplemental electronic files associated with this paper.

\subsubsection{\kic}
\label{sec:1255}

Figure \ref{fig:k4} also shows the DTS of \kic\, for Q1-Q16 on the same vertical scale as the \kfour\ and \kfive\ DTS plots. Notable in this DTS are the two quiescent periods at the beginning and end of the DTS where the measured transit depth is nearly zero, and between those two quiescent areas where the depths are highly variable. This DTS was kindly provided by Bryce Croll (2015, private communication) who describes its construction in \citet{Croll14}.  We rejected one highly negative depth in the \kic\ time series as unphysical.

\begin{figure*}
\begin{center}
\plottwo{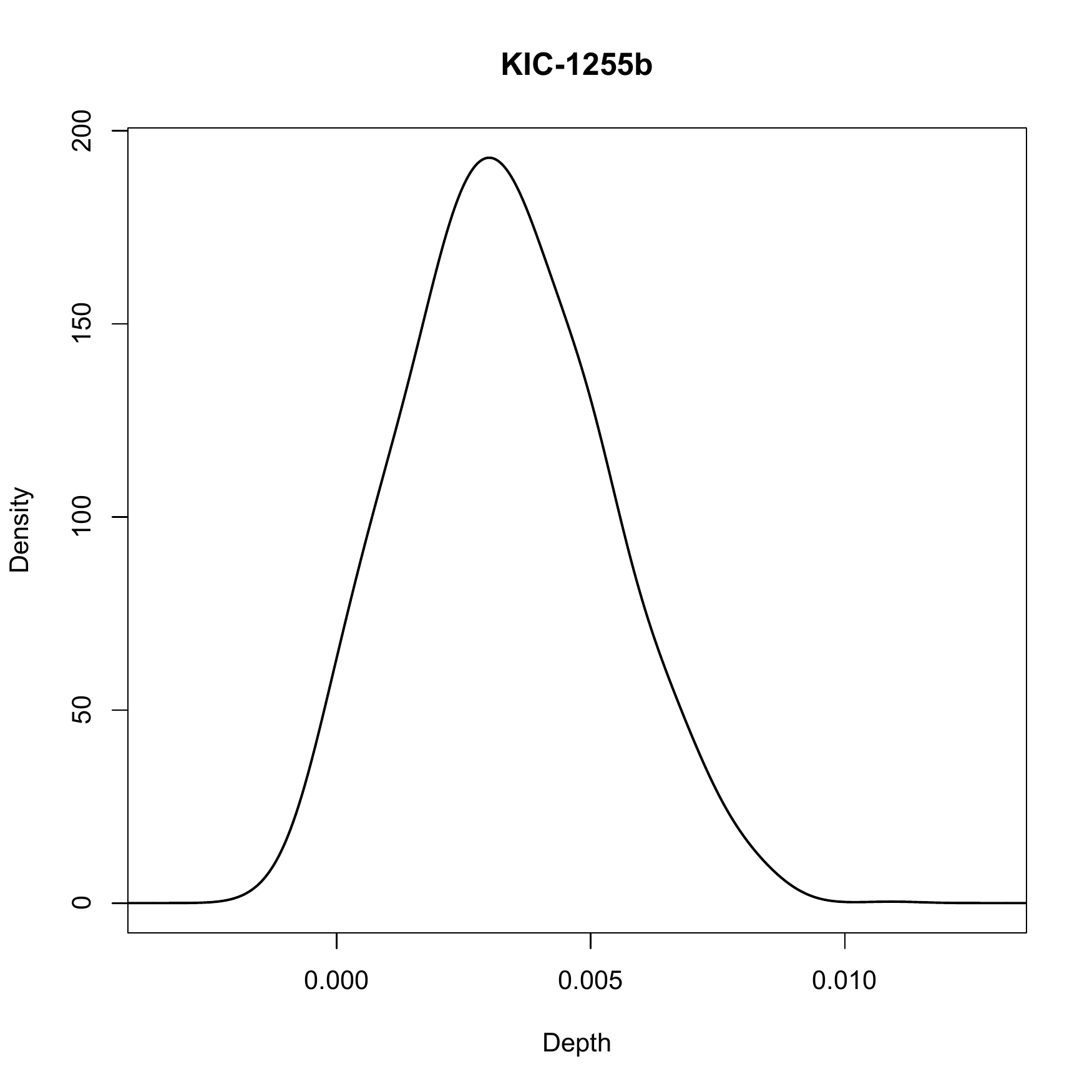}{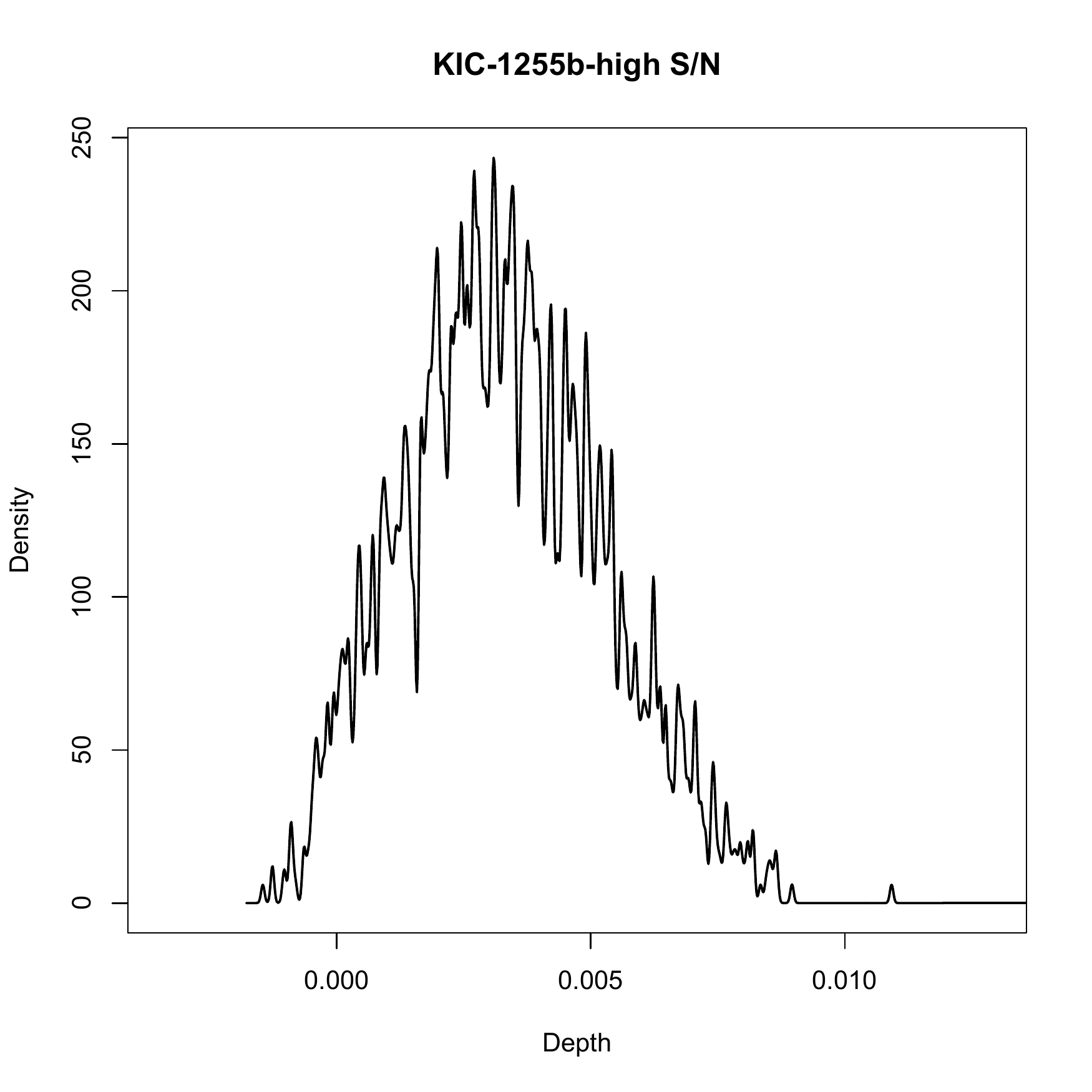}
\caption[PDF of the DTS of \kic]{PDF of the DTS of \kic\ with a convolving kernel width equal to the mean measurement uncertainty of \kic\ (``low-S/N,'' left) and \kfive\ (``high-S/N,'' right).}
\label{fig:k1255pdf}
\end{center}
\end{figure*}

Figure~\ref{fig:k1255pdf} shows the PDF for the \kic\, DTS. The PDF on the left was generated in the same manner as the \kfour\ DTS, with a kernel width equal to the average measurement uncertainty of the transit depths \citep{Croll14} The smoothness of the PDF is due to the larger width of the convolving kernel, but the width of the PDF is notably much wider than for \kfour\ and does stretch below ${\rm depth}=0$ owing to the quiescent periods. To illustrate the information content that could exist in a \kic-like system observed by {\it Kepler}, we simulated a system with the same measured depths, but at $\sim10$ times better precision (consistent with the S/N level of \kfive, labeled as ``KIC-1255b (high S/N)" in figures). The PDF generated from the convolution of the \kic\ DTS with a kernel with the width of the \kfive\, uncertainties is shown on the right of the figure.

\subsubsection{Beacon 1: \bone}

\label{sec:b1}

\begin{figure*}
\begin{center}
\plottwo{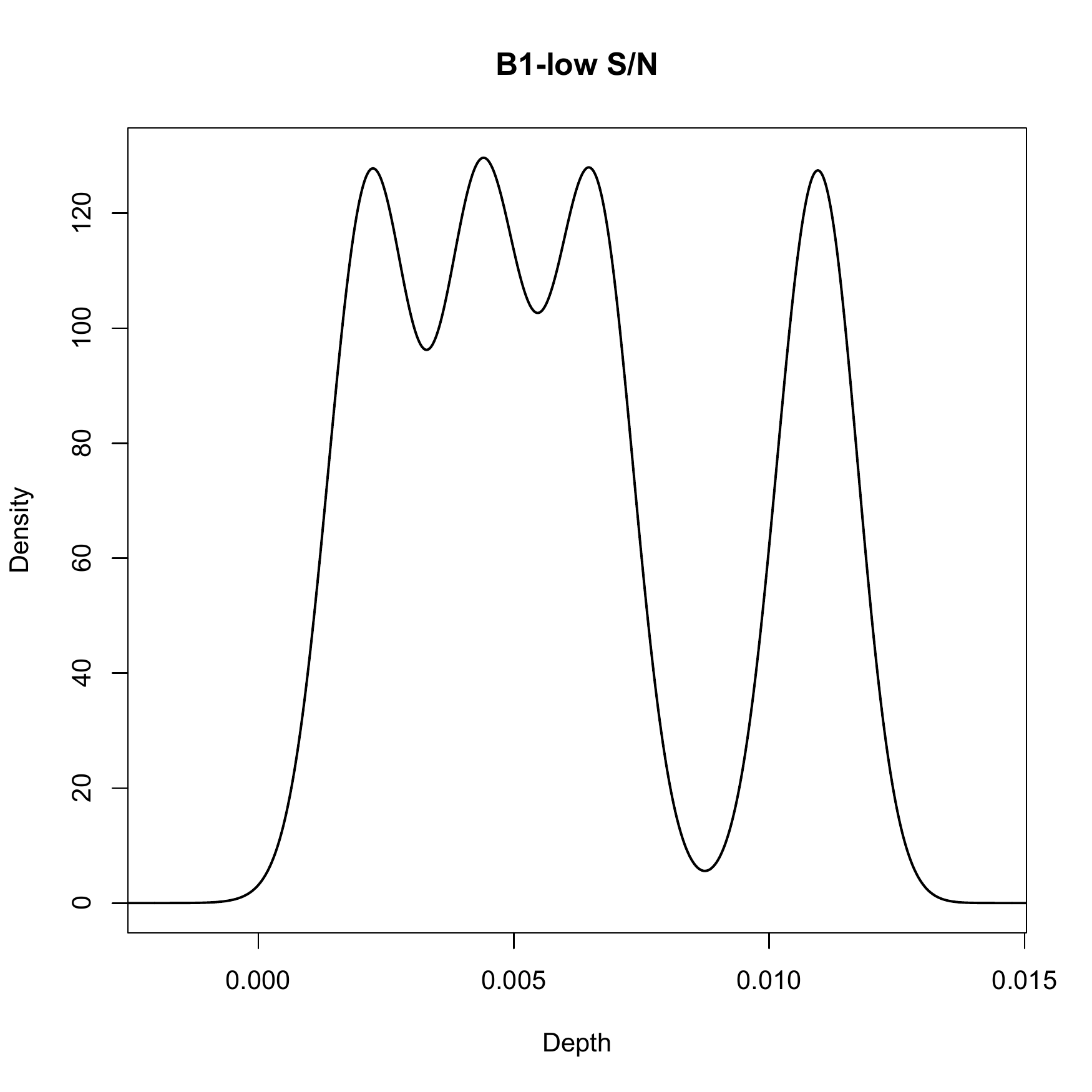}{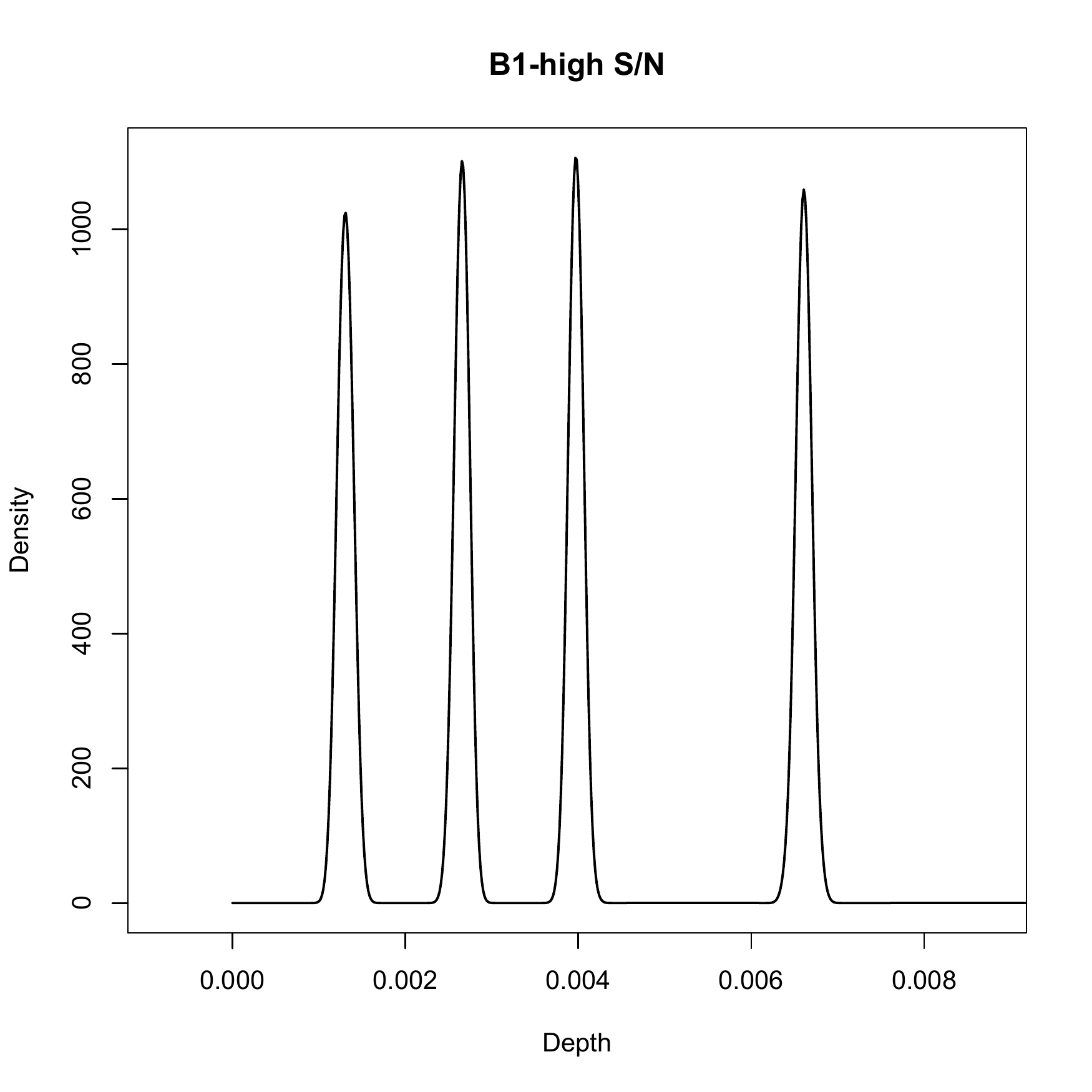}
\caption[PDF of the DTS of B1-high/low S/N]{PDFs of the DTS of B1 at low- and high-S/N, using a convolving kernel with width equal to the mean measurement uncertainty of \kic\ (left) and \kfive\ (right).}
\label{fig:b1k5pdf}
\end{center}
\end{figure*}

A simple beacon that has been considered for decades is a sequence of repeating prime numbers \citep{contact}, a simplified version of which constitutes Arnold's beacon: a series of co-orbital objects whose combined signature is a repeated series transits with depths following the pattern $\overline{\bone}$.  We test this signal at both a high S/N and a low S/N to simulate our current detection capabilities.

We generate the high-S/N DTS case for Beacon 1 (labeled as ``B1-high S/N") by repeating $\bone/5\times\max({d})$ (where max($d$) is the maximum depth of the \kfive\ signal) an integer number of times to match as closely as possible the length of the \kfive\ DTS, as would be observed by Arnold's suggestions, ignoring the gaps between the transits.  We then add to this noise at the level of the \kfive\ uncertainties by randomly sampling from the gaussian distribution $N(\mu=0, \sigma=\sigma_{d,{\rm K5}})$.

The PDF for this case is shown on the right side of Figure \ref{fig:b1k5pdf}. This was generated by convolving the B1-high S/N DTS with a Gaussian kernel with a width equal to $\sigma_{d,{\rm K5}}$. We generate the low-S/N DTS case for Beacon 1 (labeled as ``B1-low S/N") by repeating the same sequence an integer number of times to most closely match the length of the \kic\ DTS, this time adding to this noise at the level of the \kic\ uncertainties by randomly sampling from the Gaussian distribution $N(\mu=0, \sigma=\sigma_{d,{\rm K1255}})$. The PDF for this case is shown on the left side of the figure. This was generated by convolving the B1-low S/N DTS with a  gaussian kernel with a width equal to $\sigma_{d,{\rm K1255}}$.

\subsubsection{Beacon 2: \btwo}

\label{sec:b2}

\begin{figure*}
\begin{center}
\plottwo{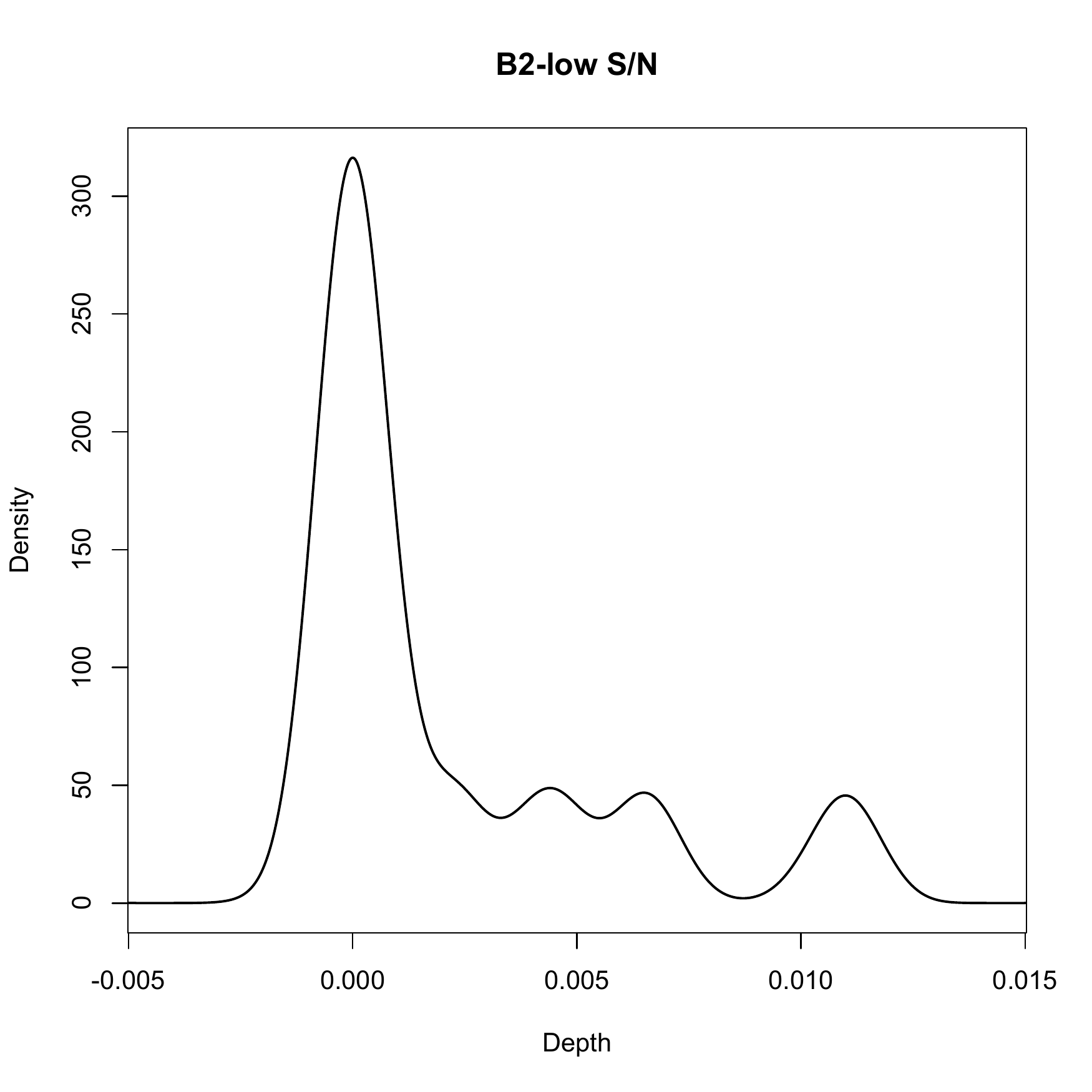}{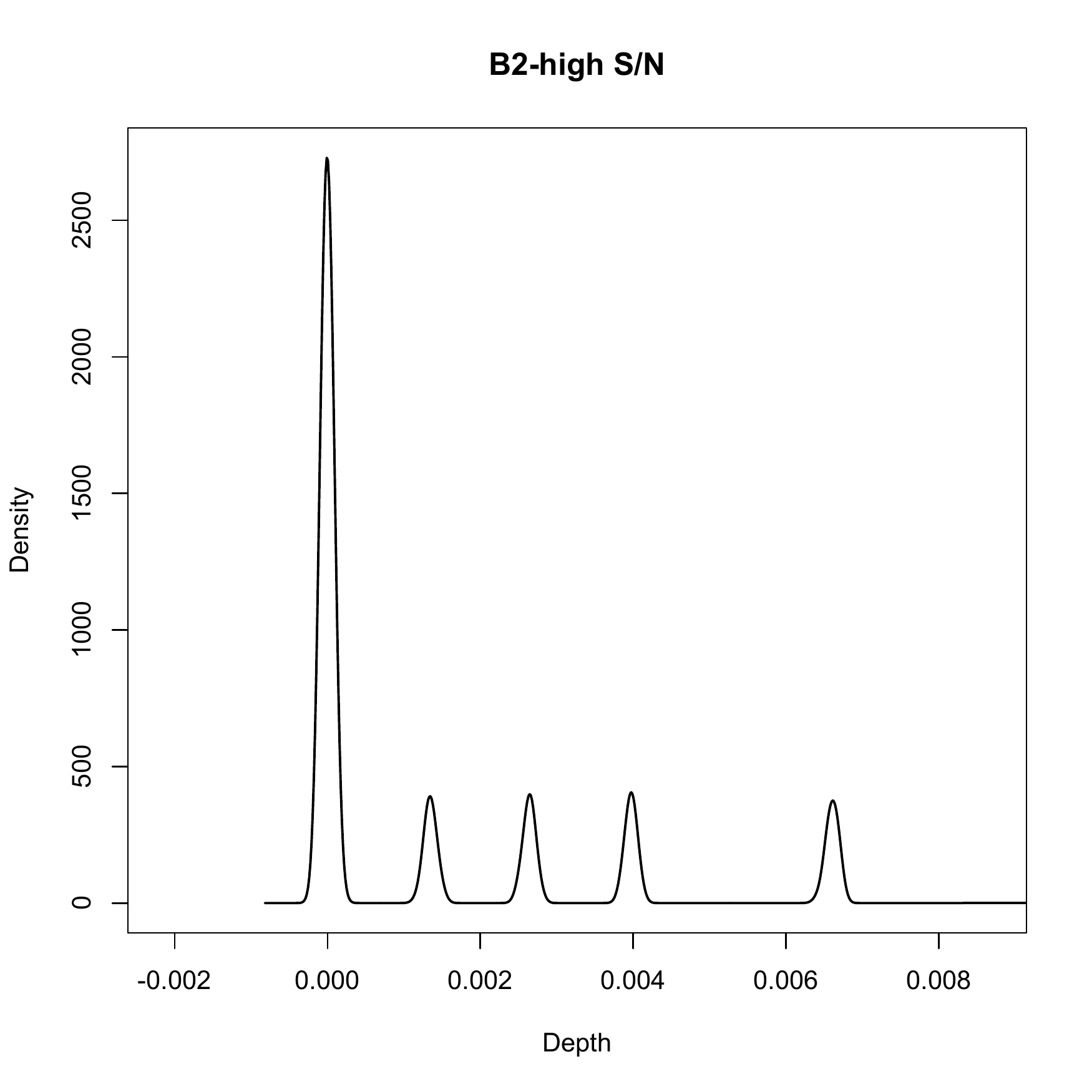}
\caption[PDF of the DTS of B2-high/low S/N]{PDF of the DTS of B2 at low- and high-S/N, using a convolving kernel with width equal to the mean measurement uncertainty of \kic\ (left) and \kfive\ (right).}
\label{fig:b2k5pdf}
\end{center}
\end{figure*}

Arnold's beacon also encoded the prime number sequence a second way: in the spacings of the transit events \citep[see Figure 8 of][]{Arnold05}.  We accommodate this with a second interpretation of the signal of Arnold's beacon, by constructing a DTS as a repeating sequence with events spaced by the narrowest gap between transits, and containing an appropriate number of null transits (depth = 0) between the more widely spaced transits: \btwo. 

We again test this beacon at both a high and low S/N to simulate our current detection capabilities. The high-S/N and low-S/N synthetic DTS for Beacon 2 were generated in the same way as for Beacon 1, as described in Sec. \ref{sec:b1}.

The PDF for the high-S/N case is shown in the right side of Figure \ref{fig:b2k5pdf}. This was generated by convolving the B2-high S/N DTS with a Gaussian kernel with a width equal to $\sigma_{d,{\rm K5}}$. The PDF for the low-S/N case is shown on the left side of the figure. This was generated by convolving the B2-low S/N DTS with a  gaussian kernel with a width equal to $\sigma_{d,{\rm K1255}}$.

\subsection{Frequency-space Analysis of Beacons and Real Transiting Systems}

\begin{figure*}
\begin{center}
\plotone{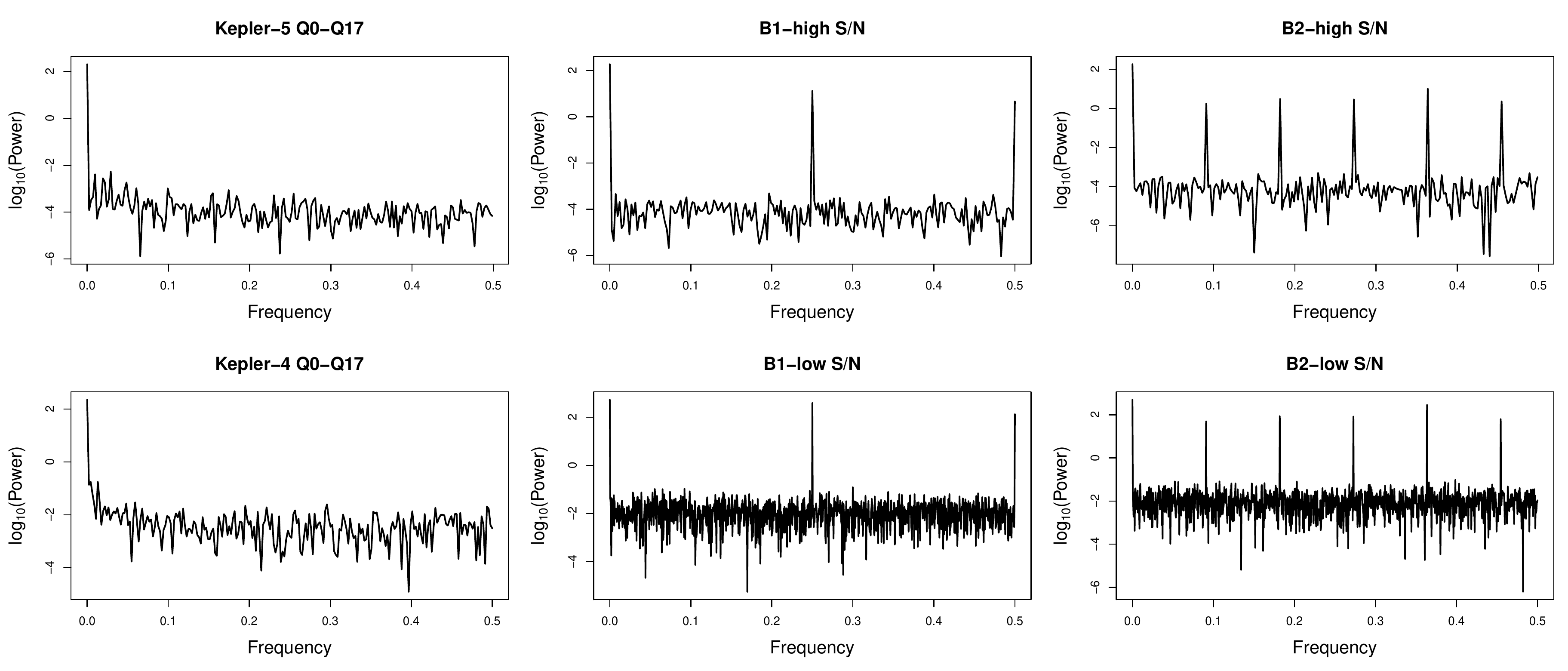}
\caption[Folded and Normalized DFTs]{Comparative normalized, folded DFTs for the six cases of \kfour, \kfive, and the beacons at high and low S/N. Frequency units are $1/p\, (day^{-1})$. Note the large variation in the scales of the (logarithmic) y-axes.  \kfour\ and \kfive\ appear consistent with constant depth plus nearly white noise.  B1 and B2 show significant power at a small number of frequencies.  Note that the normalization procedure makes the level of the noise sensitive to the amount of power in the frequencies present in the signal (see text for more detail.)
}
\label{fig:dft}
\end{center}
\end{figure*}

\begin{figure*}
\begin{center}
\plotone{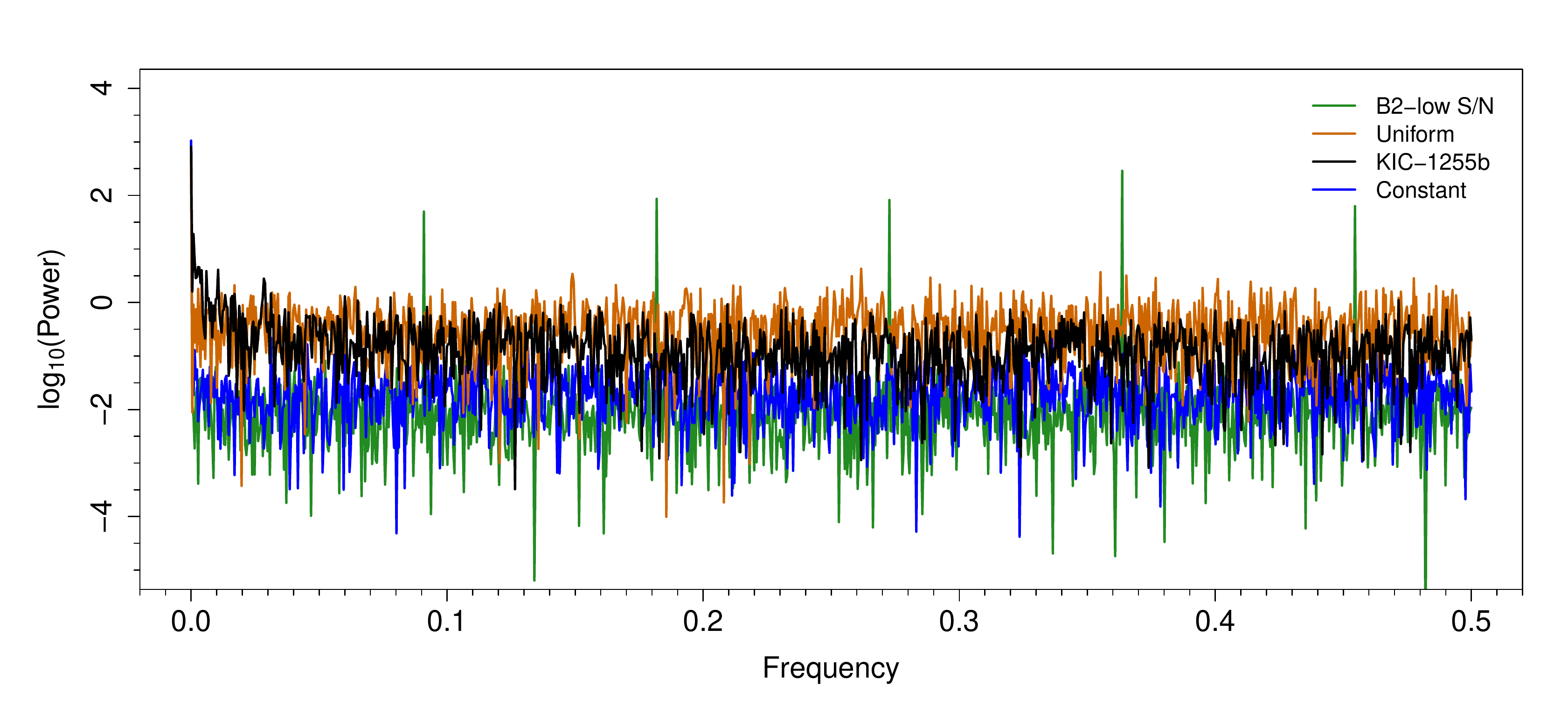}
\caption[Log-Scale DFTs]{Comparative normalized, folded power spectra for \kic, B2 at low S/N, the uniform (maximal information) case, and the constant (zero-information) case. Note the log y-axis, and that the frequency units are $1/p_{1255}\, (day^{-1})$, so the Nyquist frequency is at 0.5.  

The overall level of the power at most frequencies for \kic\ is intermediate between the constant and uniform cases, as we expect for a stochastic, but sub-maximal, signal.  \kic\ shows no obvious periodicities, except for some power at low frequencies due to the long ``quiescent'' periods at the beginning and end of the {\it Kepler} observations (see Figure~\ref{fig:k4}, bottom).  

Note how, because of the normalization procedure, the noise level for the beacon appears lower than the constant case, although the two signals actually have the same amount of noise.}
\label{fig:log}
\end{center}
\end{figure*}

We implemented the same procedure described above in order to calculate the relative information content of the folded, normalized power spectrum of the depth sequences, that is, the $M$-values in frequency space.  For the high-S/N cases of the beacons, we used the length, measurement noise, and DC term associated with \kfive, and for the low-S/N cases of the beacons we used the length and measurement noise, and DC term associated with \kic.

The Fourier transform of a time series is sensitive to the treatment of missing data.  This is important because the full depth sequences of \kfour, \kfive, and \kic\ contain gaps between the 17 ``quarters'' that defined the {\it Kepler} observing campaign, and in some cases whole quarters are missing due to module failure on the spacecraft.  In our calculations of the discrete Fourier transform (DFT) of the full sequences, we chose to linearly interpolate between the endpoints of each quarter and generate simulated depths with simulated measurement noise at each expected transit time.  There are also some missing depths within each quarter due to instrumental glitches and other sources of ``bad data'' (indicated by poor fits to the light curves, see Section~\ref{sec:kfour}).  We used the same interpolation procedure to provide artificial depths for these transits.  We include our final DTS series for our frequency analyses in our supplemental electronic files associated with this paper.

To explore how this procedure may have induced periodicity in the DFT that is not truly there, we also calculated the DFT of each available individual quarter, which provide a sense for how the $M$-value might vary with time on shorter segments of the DTS. 

Fig. \ref{fig:dft} shows the folded, normalized DFTs for the six cases explored in this section. Here the x-axis is scaled to the same range for each case to better compare the DFTs with a frequency unit of $1/{\rm orbital\ period\ (day^{-1})}$. 

The spectra for \kfour\ and \kfive\ are as expected for constant signals, although \kfour\ shows a small excess of low-frequency power above white noise that may result from our interpolation across observing gaps or systematic photometric noise in the {\it Kepler} data (\kfive\ may show a similar, smaller excess)  This will serve to distinguish these cases slightly from ideal constant cases.  B1 shows power primarily at frequencies of 0.25 and 0.5, while B2 shows power at five frequencies, due to the more complex nature of the way we have interpreted the signal.

Figure~\ref{fig:log} shows the normalized, folded DFTs for \kic, the constant and uniform (maximal) comparison cases, and the low-S/N case of B2.  As expected, the \kic\ power spectrum is consistent with noise at a level intermediate to the constant and uniform cases, though significantly closer to the uniform case.  The beacon, consistent with its nature, shows a simple structure with power at small number of discrete frequencies.  The normalization of the DFTs serves to make the level of the noise in the beacon appear lower than that in the constant case, although the time series contain the same amount of noise.

\subsection{Results of the Information Content Analysis}

\subsubsection{Results for the DTS Analysis}

\begin{figure*}
\begin{center}
\plotone{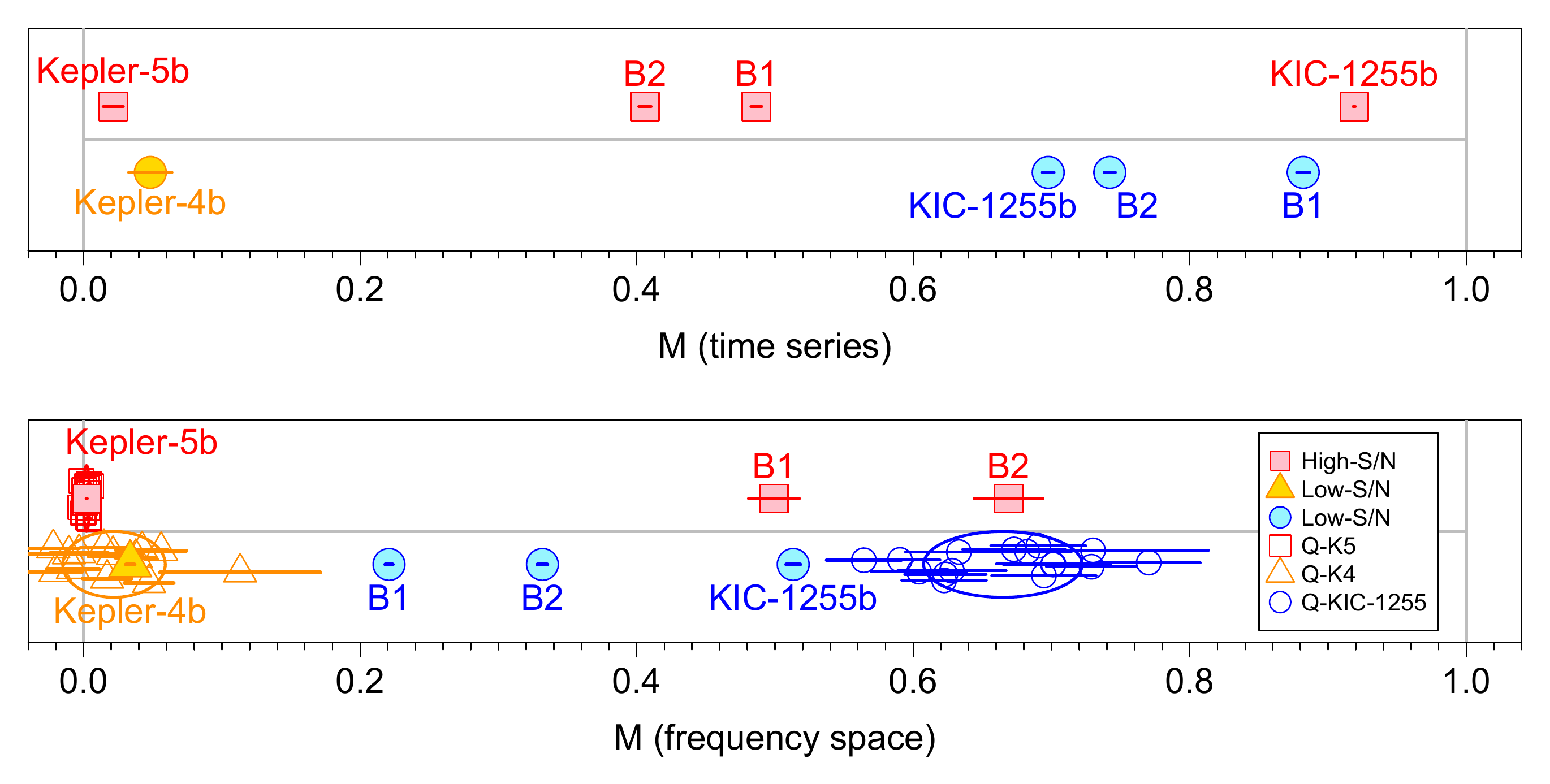}
\caption[Relative Entropy Statistic of Depth Time Series]{Normalized information content, $M$, of the depth time series data (top) and their power spectra (bottom) for various cases discussed in the text.  ``High-S/N'' cases, as red squares, refer to \kfive, the beacons observed and analyzed at the same number of transits and same S/N ratio as \kfive, and (in the top panel) a hypothetical version of \kic\ where we have treated the actual, measured time series of depths as if they were measured at this S/N.    ``Low-S/N'' cases, as blue and yellow symbols, refer to \kfour, the beacons observed and analyzed with the same number of transit events and S/N as \kic, and \kic\ itself.   In the bottom panel, individual \kepler\ quarters appear as open symbols, and the large ellipses have horizontal axes widths equal to the standard deviation of the quarters they are centered on. 

The vertical axis is not quantitative and serves only to separate the various cases for clarity.  The horizontal bars within each symbol represents the uncertainty in $M$, with important caveats described in Appendix~\ref{uncertainty}.

In all cases, \kfour\ and \kfive\ are seen to be clearly nearly information-free, and high-S/N beacons have intermediate values of $M$ in both bases, as expected given their simple structure.   

In the time domain (top), \kic\ exhibits near-maximal information content, consistent with the nearly uniform distribution of its measurements.  At lower S/N, the beacons have higher $M$ values, because the lower precision obscures the small number of values they take on, making their depth distributions more consistent with a uniform distribution.

In the frequency domain (bottom), \kic\ exhibits intermediate values for $M$, revealing significant non-random structure in the depth time series that is nonetheless significantly more complex than the simple beacons.  The much longer time series of the ``low-S/N'' case makes the $M$ values much more precise in this domain, overwhelming the effects of the lower S/N.  

}
\label{fig:mstatistic}
\end{center}
\end{figure*}

Following the procedure described above, we calculated normalized information content $M$. Table \ref{tab:depthsolution} contains the values that went into calculating Eq. \ref{eq:statistic} as well as the final statistic values for each case. Figure \ref{fig:mstatistic} (top) shows where the information content of each case falls on the statistic. 

From both of these we can see that the constant cases of \kfour\ and \kfive\ are well measured as having near-zero information content.  Their non-zero values are likely due to small systematic errors in the photometry in excess of the (very low) shot noise, which broaden the PDF of the measured depths slightly in excess of that for an ideal Gaussian with a width given by the median of the formal measurement errors.  The error bars do not encompass zero in part because we have not simulated measurement noise in our calculation of $K_m$ (as described in Section~\ref{sec:errors}).

The high-S/N cases all have S/N$\approx 100$, as the beacons would if they had been observed by {\it Kepler} and they had the depth, brightness, and transit frequency of \kfive.  We see that, as we anticipated, the beacons have intermediate information content, with $M$ values near 0.5.  The B2 case scores slightly lower because we have chosen to represent the gaps with many zeros, making the distribution of ``depths'' less uniform.  

Also as we expected, \kic\ scores very high on the $M$ statistic if we grant the measured depths false precision and assign them the very low measurement noise of \kfive.  What we are seeing here is that the measurement noise is information-rich in the sense that it spans many of the values within its range, unlike the beacons which take on only a few values.

The low-S/N cases give very different results.  Because the S/N in this case is much lower ($\sim 15$), the beacons are no longer detected as having a discrete series of depths.  Rather, they appear to span the range from $[0,\max({\rm depth})]$ rather uniformly, and so have very high $M$-values.  Interestingly, at this S/N \kic\ actually scores lower than the beacons (or itself at high S/N) because we are now more sensitive to the non-uniformities in the depth PDF (the highest depths are underrepresented).

We conclude from this that the relative entropy statistic of the DTS data is a good way to distinguish constant stars, simple beacons, and ``random'' or information-rich signals if those signals are detected at high S/N.  We also conclude that \kic\ cannot be yet be distinguished from a beacon (in its DTS) because it has not been measured at sufficient precision to exclude the possibility that the transits exhibit only a small number of discrete depths.

\begin{table*}
\caption{Depth Time Series Relative Entropy Values}
\begin{center}
\begin{tabular}{ c c c c c c c }
\hline
Case 		& max($d$) (ppm)  & $\sigma_d$ (ppm) 	&	$K_m$	&	$K_0$			&	$K_{\rm max}$		&	$M$	\\ \hline \hline
\kfive		& 6670                         & 64                         &	$7.6833$&	$7.8833\pm0.0185$	&	$5.0355\pm0.0113$	&	$0.0215\pm0.0073$\\
B1-high S/N 	& 6670                         & 64                         &	$6.4921$&	$7.8837\pm0.0188$	&	$5.0362\pm0.0117$	&	$0.4887\pm0.0039$\\
B2-high S/N	& 6670                         & 64                         &	$6.7204$&	$7.8833\pm0.0195$	&	$5.0374\pm0.0126$	&	$0.4086\pm0.0045$\\
\kic-high S/N	& 10900\tablenotemark{1}                      & 32                         &	$4.8589$&	$8.5507\pm0.0079$	&	$4.5331\pm0.0034$	&	$0.9189\pm0.0008$\\
\kfour              & 790                          & 40                         &	$8.2801$&	$8.3412\pm0.0207$	&	$7.0821\pm0.0193$	&	$0.0483\pm0.0157$\\
\kic		        &  10900\tablenotemark{1}                     &  561                      &	$4.7908$&	$5.7207\pm0.0083$	&	$4.3877\pm0.0070$	&	$0.6976\pm0.0043$\\
B1-low S/N	&  10900                      &  561                      &	$4.5466$&	$5.7204\pm0.0079$	&	$4.3876\pm0.0071$	&	$0.8806\pm0.0048$\\
B2-low S/N	&  10900                      &  561                      &	$4.7189$&	$5.7211\pm0.0080$	&	$4.3876\pm0.0067$	&	$0.7516\pm0.0041$\\ \hline
\end{tabular}
\end{center}
Notes: Values listed for $K_0$, $K_{\rm max}$, and $M$ are the mean and $1\sigma$ of the ensemble of values calculated.\\
\tablenotemark{1}{The median and 99th percentile depths for \kic\ are 3200 and 7950 ppm, giving S/N values of $\sim 100$ and 250 in the high-S/N case, and $\sim 6$ and 14 in the low-S/N case.}
\label{tab:depthsolution}
\end{table*}%

\subsubsection{Results for Frequency Space Analysis}

We present the results for the DFT calculation of the relative entropy statistic for the several cases explored in Table~\ref{tab:dftsolution}, and in Figure~\ref{fig:mstatistic} (bottom). We present the solutions for the individual quarters in Table \ref{tab:quarters}. As in our time series analysis, the constant cases show low information content, consistent with zero, and the beacons in the high-S/N case again have intermediate values of $M$, around 0.5.  

In contrast to the time-series analysis, our frequency analysis properly distinguishes \kic\ from the beacons, even at low S/N.  This makes sense because our ``low'' S/N case actually contains five times as many data points as the high-S/N case, which enhances the power of the beacons in frequency space and makes their simple structure more prominent.  The constant cases, the pseudo-random \kic\, case, and the simulated beacons thus land roughly where we expect them to. 

The individual quarters group around their full time series average very well for the \kfour\ and \kfive\ cases, as we expected.  However, for \kic\ the full series shows significantly less information constant than the average of the individual quarters --- indeed the statistic values for the individual \kic\ quarters do not encompass the value for the full depth sequence. When we examine the DTS of \kic\, (Fig. \ref{fig:k4}, bottom) we see that there are two quiescent periods at the start and end of the DTS, which adds power at low frequency in the power spectrum for the full sequence (see Figure~\ref{fig:log}) that is not apparent in the individual quarters. 

Three quarters had fewer than 10 detected transits because they were shorter than usual: quarters 0 and 4 for \kfour, and quarter 0 for \kfive. Because of the extremely low $n_d$ for these quarters, the statistic values were unreliable, and we do not include them in the plot. 

\begin{table*}
\caption{Frequency Space Relative Entropy Values}
\begin{center}
\begin{tabular}{c c c c c c }
\hline
Case 		& $n_d$&	$K_m$	&	${K_0}$		&	$K_{\rm max}$	&	$M$			\\ \hline \hline
\kfive		& 413 &	5.3296	&	$5.3312\pm0.0001$	&	$4.6411\pm0.0249$	&	$0.0023\pm0.0002$\\
B1-high S/N	& 412 &	4.9872	&	$5.3312\pm0.0001$	&	$4.6424\pm0.0246$	&	$0.5001\pm0.0179$\\
B2-high S/N	& 407 &	4.8558	&	$5.3166\pm0.0001$	&	$4.6287\pm0.0258$	&	$0.6708\pm0.0253$\\
\kfour		& 456 &	5.3512	&	$5.3843\pm0.0029$	&	$4.4028\pm0.0353$	&	$0.0337\pm0.0032$\\
\kic		        & 2182 &	4.8283	&	$6.6762\pm0.0084$	&	$3.0761\pm0.0340$	&	$0.5133\pm0.0049$\\
B1-low S/N	& 2180 &	5.8540	&	$6.6750\pm0.0087$	&	$3.0748\pm0.0335$	&	$0.2280\pm0.00294$\\
B2-low S/N	& 2178 &	5.4401	&	$6.6747\pm0.0082$	&	$3.0742\pm0.0334$	&	$0.3429\pm0.0036$\\ \hline
\end{tabular}
\end{center}
Note: Values listed for $K_0$, $K_{\rm max}$, and $M$ are the mean and $1\sigma$ of the ensemble of values calculated.  See Table~\ref{tab:depthsolution} for $\sigma_d$ and max($d$) values.
\label{tab:dftsolution}
\end{table*}%

\begin{table*}
\caption{$M$ (Frequency Space) for \kepler\, Quarters}
\begin{center}
\begin{tabular}{c c c c}
\hline
Quarter	&	\kfive			&	\kfour			&	\kic			\\ \hline \hline
Q0		&	$-0.0247\pm1.6156$	&	$0.6930\pm9.1294$	&	\nodata		        \\
Q1		&	$-0.0023\pm0.0021$	&	$-0.0092\pm0.0463$	&	$0.7296 \pm 0.0796$	\\
Q2		&	$0.0034\pm0.0010$	&	$0.0378\pm0.0158$	&	$0.6870 \pm 0351$	\\
Q3		&	$0.0029\pm0.0009$	&	$0.0562\pm0.0184$	&	$0.6729 \pm 0368$	\\
Q4		&	$0.0026\pm0.0009$	&	$0.0368\pm0.2640$	&	$0.6219 \pm 0302$	\\
Q5		&	$0.0026\pm0.0008$	&	$0.0170\pm0.0176$	&	$0.7726 \pm 0384$	\\
Q6		&	$0.0036\pm0.0010$	&	$-0.0203\pm0.0181$	&	$0.6994 \pm 0407$	\\
Q7		&	$0.0044\pm0.0011$	&	$-0.0036\pm0.0197$	&	$0.6953 \pm 0374$	\\
Q8		&	$0.0055\pm0.0012$	&	\nodata				&	$0.6315 \pm 0368$	\\
Q9		&	$0.0005\pm0.0007$	&	$0.0467\pm0.0183$	&	$0.7278 \pm 0347$	\\
Q10		&	$0.0036\pm0.0009$	&	$-0.0219\pm0.0190$	&	$0.5613 \pm 0264$	\\
Q11		&	$0.0022\pm0.0008$	&	$0.0141\pm0.0175$	&	$0.6047 \pm 0340$	\\
Q12		&	$0.0009\pm0.0008$	&	\nodata				&	$0.6211 \pm 0298$	\\
Q13		&	$-0.0003\pm0.0008$	&	$-0.0077\pm0.0188$	&	$0.6847 \pm 0309$	\\
Q14		&	$0.0042\pm0.0010$	&	$0.0210\pm0.0168$	&	$0.5896 \pm 0273$	\\
Q15		&	$0.0040\pm0.0009$	&	$0.0414\pm0.0191$	&	$0.7009 \pm 0355$	\\
Q16		&	$0.0001\pm0.0008$	&	\nodata				&	$0.6281 \pm 0365$	\\
Q17		&	$-0.0017\pm0.0018$	&	$0.1151\pm0.0583$	&	\nodata			\\ \hline
\end{tabular}
\end{center}
\label{tab:quarters}
\end{table*}


\subsection{Effects of S/N and Signal Length on the Normalized Information Content Metric}

As we have seen, the value of $M$ for a given signal can be strongly dependent on the S/N at which it is observed and the length of the signal, because the metric is normalized by the maximal information one {\it could} measure at a given S/N and signal length.  In the time domain, the S/N is determined by the measurement precision, and the value of $K_{\rm max}$ is sensitive to the signal length.

This means that at low S/N, one is not very insensitive to information content in the signal (i.e.\ $K_0$ and $K_{\rm max}$ are similar, especially for short signals).  The actual behavior of $M$ with increasing S/N depends on the nature of the underlying signal, and how information is revealed with higher precisions

This is illustrated nicely by the opposing behaviors of \kic\ and the Arnold Beacons in Fig.~\ref{fig:mstatistic}: at high S/N, the Arnold beacons clearly take on only a few discrete values (low information content). But because these peaks are roughly evenly distributed about the span of the depths, at lower precision the discreteness is lost and the values appear uniformly distributed, and the normalized information content goes up.  In a hypothetical case of much more tightly spaced beacon values (say, a hundred discrete values between 0.10 and 0.11), the opposite effect might occur: at low S/N there might appear to be a single, narrow, well-defined peak (low $M$), but at high S/N the shallower peak might resolve into a large number of discrete peaks, and thus show very high information content.  Indeed, this is similar to the behavior of \kic\ in the time domain, which has a preferred range of depths that appears to break into many discrete peaks at high S/N.

In the frequency domain, increasing the length of a time series strongly increases one's sensitivity to strictly periodic components of a signal.  In our analysis of the Arnold beacons, this effect dominates over the photometric precision improvement (recall that our ``low-S/N'' test case is \kic\ which, having a shorter period than \kfive, has many more depth points).  As a result, the Arnold beacons have much smaller error bars in the ``low-S/N'' case, and lie nearer the values they would have in an infinitely long signal.
 
Use of the $M$ statistic thus requires comparison to nominal signals {\it at the same signal length and S/N as the signal in question}.  In the case of \kic\, we can conclude that we have measured much more information than a constant signal or our beacons in frequency space, which is consistent with the signal being complex but not maximally random. In the time domain we measure a very high information content (as expected from a complex signal) but, because our S/N is low, this is similar to the result we get for our beacons in the time domain, showing that our precision does not give us sensitivity to very complex signals.

\section{Conclusions}

Arnold (2005), Forgan (2013), and Korpela et al.\ (2015) noted that planet-sized artificial structures transiting their host star could be discovered with {\it Kepler}.  They noted that such structures might be used for stellar energy collection, propulsion, or as beacons, providing very long-lived, luminous modes of interstellar communication with small marginal cost per bit.  

Invoking alien engineering to explain an anomalous astronomical phenomenon can be a perilous approach to science because it can lead to an ``aliens of the gaps'' fallacy \citep[as discussed in \S 2.3 of][]{GHAT1} and unfalsifiable hypotheses. The conservative approach is therefore to initially ascribe all anomalies to natural sources, and only entertain the ETI hypothesis in cases where even the most contrived natural explanations fail to adequately explain the data.  Nonetheless, invoking the ETI hypothesis can be a perfectly reasonable way to enrich the search space of communication SETI efforts with extraordinary targets, even while natural explanations are pursued \citep{GHAT2,Teodorani2014,Bradbury2011}. 
 
To that end, and without committing to a particular form of or purpose for megastructures, we have enumerated ten potential ways their anomalous silhouettes, orbits, and transmission properties would distinguish them from exoplanets.  Many of these signatures mimic transit anomalies caused for natural reasons, but some would be very difficult to explain naturally. {\it Kepler} thus has the potential to detect or put tight upper limits on the frequency of structures above certain sizes in short period orbits around its target stars via (non-)detection of these signatures.  We recommend that a future search for alien megastructures in photometric data sets such as that of {\it Kepler} search for these ten signatures.  Objects exhibiting more than one of these signatures should be especially scrutinized.
 
Since predictions often carry more rhetorical and philosophical weight in science than post-hoc explanations \citep[e.g.,][]{predictions} we believe it is worthy of citation that Arnold predicted that {\it Kepler} might detect transit signatures similar in many ways to those seen in KIC 12557548{\it b} and CoRoT-29{\it b}, and presaged \wtf\ in some ways, as well. We note that several other anomalous objects, too, have variability in depth consistent with Arnold's prediction, and/or an asymmetric shape consistent with Forgan's model. Since evaporating-planet models of \kic\ by Rappaport et al. (2012) have so far provided a satisfactory and natural explanation for \kic, and radial velocities appear to confirm that CoRoT-29{\it b} has planetary mass, the ETI hypothesis for these objects is not warranted at this time.  But these objects can still serve as useful examples of how non-standard transit signatures might be identified and interpreted in a SETI context.
 
A comprehensive search for megastructures should consult the original light curves, since the standard {\it Kepler} transit-detection and assessment pipelines are not ``looking for'' megastructures that may be present in the data --- that is, the frequencies of anomalous transits are not naturally computed as part of normal {\it Kepler} planet frequency statistics, both because ordinary {\it Kepler} data analysis is not sensitive to many megastructure transit signatures, and because such signatures may be mistakenly ascribed to natural sources.  Indeed, in some cases of highly non-standard transit signatures, it may be that only a model-free approach --- such as a human-based, star-by-star light curve examination --- would turn them up.  Indeed, \wtf\ was discovered in exactly this manner.  \wtf\ shows transit signatures consistent with a swarm of artificial objects, and we strongly encourage intense SETI efforts on it, in addition to conventional astronomical efforts to find more such objects (since, if it is natural, it is both very interesting in its own right and unlikely to be unique).  Since the {\it Kepler} data archive presumably contains many poorly studied stars that may exhibit these signatures, the alien megastructure rate remains poorly constrained.

We have developed the {\it normalized information content} statistic $M$ to quantify the information content in a signal embedded in a discrete series of bounded measurements, such as variable transit depths, and show that it can be used to distinguish among constant sources (i.e.\ those with zero information content), interstellar beacons (having small but non-zero information content), and naturally stochastic or artificial, information-rich signals.  We have developed a treatment for $M$ in both the time and frequency domains, noting that a signal can be a beacon in one and information-rich in the other.  We have also shown how the measurement of $M$ is affected by measurement uncertainties, and (in the frequency domain) the length of the signal being analyzed.  

We have applied this formalism to real {\it Kepler} targets and a specific form of beacon suggested by Arnold to illustrate its utility.  We have used \kic\  as an example of a stochastic signal, our stand-in for a beacon or an artificial, information rich signal; \kfour\ as a constant source measured at similar S/N as \kic; and \kfive\ as a constant source measured at high S/N.  We have shown that in the time domain, the measurement uncertainties for \kic\ are too large to distinguish the signal we see from a beacon (that is, we cannot determine whether the spectrum of depths is continuous or composed of a small number of discrete depths).  In the frequency domain, however, the system shows no significant periodic structure, and is easily distinguished from simple beacons.  

{\it Facility:} \facility{Kepler}

\acknowledgments

We thank Bryce Croll for sending us time series depth data for KIC 12557548, and Tim van Kerkhoven for discussing the details of their careful time series analysis of the data.  We are indebted to Tabetha Boyajian and Saul Rappaport for sharing their data on \wtf\ with us prior to publication of their discovery paper, and for many stimulating discussions on its possible nature.  We thank Andrew Siemion and Tabetha Boyajian for discussions about \wtf\ as a SETI target and for some of the language in this paper, which is adapted from a telescope proposal they helped write.

We thank Steinn Sigur\dh sson, Thomas Beatty, Ben Nelson, Sharon Wang, and Jason Curtis for helpful discussions, and the anonymous referee for their helpful comments.

A rough outline of the authors' primary contributions to this paper follows:  J.T.W.\ conceived of and initiated this research, and composed the bulk of this paper.  K.M.S.C.\ performed the analysis, the figure preparation, and the writing of much of the text in Section~\ref{sec:beacons} and Appendix~\ref{appendixB}.  M.Z.\ contributed text and expertise to Section~\ref{sec:evaporation}, and D.J-H.\ contributed text and expertise to Section~\ref{sec:features} and Appendix~\ref{photothrust}. E.B.F.\ contributed to the theoretical basis for the analysis of information content in Section~\ref{sec:beacons}. All authors contributed overall comments and corrections.

The Center for Exoplanets and Habitable Worlds is supported by the Pennsylvania State University, the Eberly College of Science, and the Pennsylvania Space Grant Consortium.

E.B.F.\ acknowledges NExSS funding via NASA Exoplanet Research Program award \#NNX15AE21G; and M.Z.\ and J.T.W.\ acknowledge NExSS funding via NASA Origins of Solar Systems award \#NNX14AD22G.
 
This research has made use of NASA's Astrophysics Data System. Some of the data presented in this paper were obtained from the Mikulski Archive for Space Telescopes (MAST). STScI is operated by the Association of Universities for Research in Astronomy, Inc., under NASA contract NAS5-26555. Support for MAST for non-HST data is provided by the NASA Office of Space Science via grant NNX09AF08G and by other grants and contracts. This paper includes data collected by the {\it Kepler} mission, funding for which is provided by the NASA Science Mission directorate.

The results reported herein benefitted from collaborations within NASA's Nexus for Exoplanet System Science (NExSS) research collaboration network sponsored by NASA's Science Mission Directorate.


\appendix 
\section{Effect of the Photo-Thrust Effect on Asterodensity Profiling}
\label{photothrust}

We can estimate the asterodensity profiling effects from the photo-thrust effect by considering the simple case of a planet on a circular orbit.  In this case, the inferred density of a star $\rho_*$ can be approximated from parameters of a fit to a transit light curve as 

\[
\rho_* \approx \frac{3\pi}{GP^2} \left(\frac{a}{R_*}\right)^3
\]

\noindent where $P$ is the period of a transiting planet in a circular orbit, $(a/R_*)$ is the ratio of the planet's semimajor axis to the stellar radius, $G$ and $\pi$ are the usual constants.  A planet on a Keplerian orbit (that is, for which there are no significant accelerations beyond that from the gravity of the host star) obeys Newton's version of Kepler's Third Law, $M_*P^2=a^3$ (in units of solar masses, years, and AU).  

We now consider a thrust as a force measured in units of the gravitational force due to the host star, 

\[
\beta = \frac{F_{\rm thrust}}{F_{\rm Kep}}
\]

Since large thrusts can produce almost arbitrary changes to an object's orbit, we will limit our discussion to the effects of small thrusts, that is those for which $\beta \ll 1$.  In the case of a purely radial thrust, the effect on an object with measured period $P$ is to reduce the acceleration due to of gravity by a factor of $(1-\beta)$, resulting in a simple modification to Kepler's Third Law:

\[
M_*(1-\beta)P^2=a_\beta^3
\]

\noindent where we have used the $\beta$ subscript to distinguish quantities in the case where $|\beta| > 0$.  The inferred density of the star is then altered such that

\[
\frac{\rho_\beta}{\rho_*} = \frac{a_\beta^3}{a^3} = 1-\beta
\]

We see that for small radial accelerations, the effect is linear with slope $-1$: the inferred density will be too low by a fraction $\beta$.  One's sensitivity to the photo-thrust effect is then proportional to the precision with which one can measure a star's density and inversely proportional to the gravitational acceleration of the object by the star.  

To give an order-of-magnitude estimate of the surface density of a structure affected by radiation pressure that could be detected by this method, we consider a thin sheet of area $A$ and surface density $\sigma$ in a circular orbit around the star of luminosity $L_*$ at semimajor axis $a$, rotating such that it is kept normal to the incoming radiation of the star throughout its orbit.  If the collector absorbs all of the incoming radiation, then we have $F_{\rm thrust}=AL_*/(4\pi c a^2)$, and $F_{\rm Kep}=GM_*A\sigma/a^2$, so 
\[
\beta_{\rm collector}=\frac{L_*}{4\pi G c M_* \sigma} = 0.8 (L_*/L_\odot) (M_*/M_\odot)^{-1} (\sigma/({\rm g\ m}^{-2}))^{-1}
\]

If we optimistically assume that stellar densities can be independently measured to a precision of $\beta \approx 1\%$, then the photothrust effect has a detectable asterodensity profile signature for megastructures around Sun-like stars with $\sigma \approx 80$ g m$^{-2}$.  Any structure with a lower surface density than this would have an easily detected photo-thrust effect, although the signal would be complicated by any nonradial direction of the thrust.  For comparison, this is around 1/5 the surface density of household aluminum foil, comparable to many thin industrial metal foils, and two orders of magnitude larger than the thinnest gold leaf.  A Jupiter-sized aluminum disk at this surface density would require some 10$^{15}$g of aluminum, which is 6 orders of magnitude less than is present in the Earth's crust.
 
Detectable thrust generated by a rocket effect requires either extremely high exhaust velocities or very short lifetimes.  If a megastructure of mass $m$  consumes propellant over a characteristic time $\tau$ (=$m/\dot{m})$ at non-relativistic exhaust velocity $v_{\rm exhaust}$, then we have

\[
\beta_{\rm rocket} \sim \frac{v_{\rm exhaust}/\tau}{GM_*/a^2} = 5.3\times10^{-3} (v_{\rm exhaust} /({\rm km/s})) (a/{\rm AU})^2 (M_*/M_\odot)^{-1} (\tau/{\rm yr})^{-1}
\]

Structures thus need very fast refueling rates (ingesting their own mass in propellant in $\sim$ years), very short lifetimes (years), or very high exhaust velocities (modern ion drives achieve $\sim$ 30 km s$^{-2}$) to have rocket thrusts detectable via asterodensity profiling.

Of course, advanced civilizations might have use other forms of thrust, including exploitation of the stellar magnetic field or wind, but the above examples help put the plausibility of detecting this effect in perspective.

\section{Notes on M}
\label{appendixB}
\subsection{Kullback-Leibler Divergence: $K$}
\label{sec:kl}
The Kullback-Leibler Divergence (KL divergence), also known as the relative entropy, estimates the amount of information lost when one probability distribution is used to approximate another. The KL divergence of a continuous probability distribution is given by

\begin{equation}
	K = \int_{-\infty}^\infty p(x)\times\ln\left(\frac{p(x)}{q(x)}\right)\mathrm{d} x
\label{eq:kld}
\end{equation}

\noindent were $p(x)$ is the PDF of the signal and $q(x)$ is the probability distribution used to approximate the signal. (Note that the integrand in Equation~\ref{eq:kld} evaluates to zero for values of $x$ such that $p(x)\rightarrow 0$).
A signal with high information content will take on many values, while a signal with no information content takes on exactly one value. So, a ``maximal" signal is represented by a uniform distribution 
and a ``minimal" signal is represented by a delta function at the mean value of the data $p(x) = \delta(x-\mu)$.

For our purposes we wish to quantify the information lost when our signal is approximated by a uniform probability distribution. This is a measure of the information content of the signal relative to its maximal value. We thus choose a normalized, uniform comparison distribution $q(x) = {\rm constant}$ for all possible values $x$ of the signal. The KL divergence of a maximal signal (with distribution $p(x)$) will then be zero, indicating that there is no difference between the signal and the uniform distribution, and a minimal (empty) signal with no measurement error will be infinite (since the integrand of Eq.~\ref{eq:kld} diverges for $x=\mu$).

We note that the KL divergence is not symmetric: exchanging $p(x)$ and $q(x)$ does not preserve the value of $K$. We have chosen $p(x)$ and $q(x)$ as we have because the alternative --- using minimal signals as our comparison --- produces infinite KL divergences (where the comparison signal has zero probability, the formula diverges) which we cannot compare. 

Since we will compute the KL divergence numerically, we must approximate this integral with a discrete sum.  This will also allow us to naturally apply the KL divergence to distributions in frequency space, computed using discrete Fourier transforms (DFTs).  In this approximation, Eq. \ref{eq:kld} becomes

\begin{equation}
	K = \frac{R}{N} \sum_{i = 1}^{N} p(x_i)\times \ln \left(R\, p(x_i)\right)
\label{eq:kl1}
\end{equation}

\noindent where $R$ is the domain over which the PDF is sampled (so $q(x) \equiv 1/R$), N is the number of bins along the PDF, and $p(x_i)$ is the probability associated with events having values in the bin centered on $x_i$. The fraction $R/N$ assumes that all of the bins in the PDF are of equal width. If this is not the case, $R/N$ is removed from in front of the sum and replaced by the variable bin width $\Delta x_i$ within the sum.  

The sum in Eq. \ref{eq:kl1} is over the entire allowed range of the signal, and the summand evaluates to zero for values of $x_i$ such that $p(x_i)\rightarrow 0$.

\subsubsection{KL Divergences for Transit DTS}

Our instant application of the KL Divergence will be to Arnold beacons of Sections~\ref{sec:b1}--\ref{sec:b2}, and to the transit DTS of ordinary exoplanets (\kfour\ and \kfive) and one with variable transit depths (\kic).  For these purposes we choose $R=1$ and define our PDFs on the domain $[0,\ 1]$ in order to encompass the physically meaningful range of possible transit depths. Applications to other signals may use different values for $R$.

We approximate our signal PDFs via KDE of a series of measurements. This procedure convolves the distribution of transit depths with a Gaussian kernel with width equal to the typical depth measurement uncertainty.  This produces a continuous distribution, and removes the problems of false precision that come with binning small numbers of points more finely than measurement precision warrants.  

The resulting PDF can extend outside the defined range $R=[0,1]$, both because the kernel widens the distribution and because real measurement noise can yield negative transit depths.  Simply rejecting the parts of the PDF outside our range leads to problems with normalization and introduces artificial features in the PDF that can yield misleading measures of the information content.  Since the signals we will consider here never have values near 1 (total obscuration) and our precision is high, we simply shift our depth measurements by a small constant to ensure that the resulting KDE falls entirely within our range $R$.  A more robust approach would redefine our KL divergence to account for the wings of our PDF outside of our range due to measurement noise and our KDE procedure, but for high precision measurements the effects on the resulting statistic will be small, so we adopt this simpler procedure here for illustrative purposes.

\subsubsection{KL Divergences in Frequency Space}
\label{sec:klfreq}
There are many ways in which information might be encoded in a signal conveyed as a discrete sequence within a specified range, and our analysis of the time series only explores one way in which the signal might be simple or complex.  For instance, while the KL divergence for a DTS as described above can measure how variable or discrete a sequence of transit depths may be, it does not distinguish between simple, repeating signals and stochastic signals that take on only discrete values. 

For example, when using a joint distribution $q(\vec{x})$ that factorizes as $\prod_{i=1}^n q(d_i)$, as in the previous section, the sequence [1, 2, 3, 1, 2, 3, 1, 2, 3, \dots] will apparently have the same relative information as the sequence [1, 3, 2, 2, 1, 2, 3, 3, 1, \dots], even though the first is strictly repeating and the second is not.  Therefore, we extend our metric into frequency space.  This corresponds to an alternative choice of $q(\vec{x})$ which is not separable in terms of $d_i$, but is separable in terms of the DFT coefficients $\vec{c}$, i.e., $q(\vec{x}) = \prod_{i=1}^n q(c_i)$.  

Because we are dealing with real-valued signals, we elect to fold our DFTs by using only the amplitudes of the positive frequencies to calculate the relative entropy. To account for the power contained within the negative frequencies, we double the power contained in the positive frequencies (except that for sequences with an even number of elements, we did not double the power at the Nyquist frequency, which has no negative counterpart).  We then rescale the frequency axis to have domain [0, 1] (so, $R=1$; note that in Figure~\ref{fig:log} we show these power spectra in physical units of frequency, not from [0, 1]).

We would like to use this folded DFT as $p(x)$ (in place of the PDF of the DTS) in our calculation of Eq. \ref{eq:kl1}, however there is a complication, because $p(x)$ must be normalized to have unit area.  The value for $K$ in frequency space will thus be very sensitive to one's choice for the value of the DC (zero frequency) term.  For instance, real constant signals measured with some amount of measurement uncertainty will have a power spectrum consistent with white noise, with typical amplitude determined by the precision of the measurement.  The normalization procedure will rescale this white noise so that the entire DFT has unit area, making the final values at most frequencies dependent on the value of the DC term and the number of frequencies sampled.

Thus, when comparing values of frequency-space $K$ from different signals (as we will do when we compute our normalized information content $M$) it is important that the DFTs of the signals have identical DC terms.  We choose the mean transit depth of measured signal (squared, since we are using power spectra).

\subsubsection{KL Divergences in Other Bases}

Our purpose in this section is to provide a quantitative, statistical description of ``beacons'' in a SETI context.  In our development of the normalized information content, we have considered only bases in the time and frequency domains, since those are the ones in which most of the beacons proposed in the literature are simple and obviously artificial.   Of course, an alien (or human) signal might encode information in some other basis than we have considered here, making neither the time nor frequency domains the appropriate ones for our information-content analysis.  Applications of the KL divergence to many other domains can be developed straightforwardly, and at any rate the two we have developed here will suffice to illustrate their utility and dependence on some of the properties of a received signal. 

\subsection{Normalized Information Content: $M$}
\label{sec:m}

\subsubsection{Effects of Measurement Noise on Ideal Values of $K$}

We wish to normalize the KL divergence $K$ onto a scale with range [0, 1], spanning the zero and maximal information cases.  We thus define the KL divergence for a maximal distribution $K_{\rm max}$; the divergence for a constant signal $K_0$; and the divergence of a real, measured signal $K_m$.  According to our formalism, ideal signals with no measurement noise will have $K_{\rm max, ideal}=0$ and $K_{0, {\rm ideal}} = \infty$.

However random measurement noise and finite signal lengths will alter the information content of any real signal.  This and the details of the construction of the underlying PDF (for instance, the KDE width) make the measured values of $K_{\rm max}$ and $K_0$ for maximal and empty signals non-zero and finite. 

This allows us to rescale $K_m$ from the values of $K_0$ to $K_{\rm max}$ that one {\it would} calculate from empty and maximal signals {\it at the same S/N and same signal length} as the measured signal. The normalized information content will thus be a function of the S/N of the signal.

\subsubsection{Calculating $K_0$ and $K_{\rm max}$ in the Presence of Measurement Noise}

The numerical values of $K_0$ and $K_{\rm max}$ in the presence of measurement noise depends on three quantities: the length of the measurement series, the precision of the measurements, and the range the values of those measurements can take.  To compute them, we match these values to the equivalent values of the signal we used to compute $K_m$.  

For instance, consider a real signal consisting of $n_d$ discrete measurements $d_i$, having mean $\mu_d$, maximum $\max(d)$, measured with precision $\sigma_d$.  We compute $K_0$ from an artificial, constant signal as $n_d$ random values drawn from a normal distribution $N(\mu_d,\sigma_d)$; and we compute $K_{\rm max}$ from $n_d$ random values drawn from a uniform distribution with Gaussian noise, $U(0,\max(d))+N(0,\sigma_d)$.

For the computation of $K$ in the time domain, we apply a Gaussian kernel with width $\sigma_d$ to the distributions of values from all three signals to construct a continuous distribution, $p(x)$, and numerically compute the KL divergence via Eq.~\ref{eq:kl1} (using $2^{16}$ points to ensure that the function is well sampled).  

In the frequency domain, we fold the DFT of each series as described in Section~\ref{sec:klfreq} and set the DC term in the in all three cases to $\mu_d^2$ (since we are using the power spectrum).  We then normalize this function to unit area (i.e.\ the sum of the terms of the folded DFT will be the number of elements in it, which, since we have folded the DFT, is $(n_d+1)/2$).  This folded, DC-corrected, normalized DFT is our $p(x)$ for Eq.~\ref{eq:kl1}.

\subsubsection{Scaling $K_m$ to Compute $M$}

Since the KL divergence gives the relative entropy of a signal, the entropy of the measured signal compared to what we would measure from an empty signal is

\begin{equation}
	\Delta S = K_m - K_0
\label{eq:ks}
\end{equation}

\noindent If $\Delta S\approx0$ (i.e.\ there is no difference between our divergences), then our signal is consistent with pure measurement noise, and we can detect no other source of information in our data.  The maximum value of $\Delta S$ is that from a maximal distribution, i.e.

\begin{equation}
	\Delta S_{\rm max} = K_{\rm max}-K_0
\label{eq:smax}
\end{equation}

We then normalize $\Delta S$ to its maximal possible value, which allows us to construct our normalized statistic of information content, $M$, on a scale from [0, 1], as we desired:

\begin{equation}
	M = \frac{\Delta S}{\Delta S_{\rm max}} = \frac{K_m - K_0}{K_{\rm max}-K_0}
\label{eq:statistic}
\end{equation}

\subsubsection{Uncertainty in the Statistics}

\label{uncertainty}

These are several sources of uncertainty in our information-content statistics.

The first is that measurement noise itself contributes some amount of entropy to the signal.  We have adjusted for this to first order by comparing the $K_m$ value we calculate with constant and uniform cases that include noise in the statistic $M$.  The second is that the noise in the frequency-space case depends strongly on the length of the signal (the number of events observed).  We account for this by measuring $K_0$ and $K_{\rm max}$ using the same number of points as $K_m$.  But different realizations of the noise will lead to different values for $K_m, K_0, K_{\rm max}$, and therefore $M$.  We account for these effects of noise on $K_0$ and $K_{\rm max}$ by recalculating these quantities for 1000 draws of the Gaussian noise. The ensemble of values for $K_0$,$K_{\rm max}$, and $M$ that we calculated from these draws give us uncertainties on these statistics.

The effects of noise on $K_m$ cannot be robustly calculated without knowledge of the underlying signal, which we cannot assume one has. This is related to the third source of uncertainty, which is that we may have only measured a small portion of the signal, and that other parts of the signal may have a different information content.  For both reasons, we frame the problem as that of measuring the information content of the portion of the signal we have actually measured, understanding that if we repeated the measurement we would get a (perhaps slightly) different number, both because of measurement noise and because the underlying signal would be different.  

\label{sec:errors}

\end{document}